\documentclass[journal]{IEEEtran}

\usepackage{cite}
\usepackage{amsmath,amssymb,amsfonts}
\usepackage{algorithmic}
\usepackage{graphicx}
\usepackage{textcomp}
\usepackage{xcolor}
\usepackage{setspace}
\usepackage{hyperref}
\usepackage{comment}
\usepackage{array}% http://ctan.org/pkg/array
\def\BibTeX{{\rm B\kern-.05em{\sc i\kern-.025em b}\kern-.08em
    T\kern-.1667em\lower.7ex\hbox{E}\kern-.125emX}}
    
\usepackage[]{subcaption}

\usepackage{booktabs} % For formal tables
\renewcommand{\paragraph}[1]{\vspace{0.3\baselineskip}\noindent\textbf{#1}}
\usepackage{pgfplots, tikz}
\usepackage{pgfplotstable}
\usetikzlibrary{pgfplots.groupplots}
\usetikzlibrary{shapes.geometric}
\usetikzlibrary{positioning,fit,shapes.geometric,backgrounds, calc}
\usetikzlibrary{arrows,decorations.markings}
\usetikzlibrary{shapes.arrows}
\usetikzlibrary{patterns}
\tikzset{textnode/.style={inner sep=0pt,outer sep=0,execute at begin node={\strut}}}
\tikzstyle{state} = [textnode,circle, draw, inner sep=0pt, outer sep=0]

\usepackage{helvet}
\pgfplotsset{
  tick label style = {font=\scriptsize\sffamily},
  every axis label = {font=\scriptsize\sffamily},
  legend style = {font=\footnotesize\sffamily},
  title style = {font=\footnotesize\sffamily},
  label style = {font=\footnotesize\sffamily}
}

\pgfplotsset{compat=1.11,
    /pgfplots/ybar legend/.style={
    /pgfplots/legend image code/.code={%
       \draw[##1,/tikz/.cd,yshift=-0.25em]
        (0cm,0cm) rectangle (3pt,0.8em);},
   },
}

\selectcolormodel{cmyk}

\usepackage{mathtools}

\usepackage{multirow}
\usepackage{graphicx}
\usepackage{listings}
\usepackage{enumitem}

\newenvironment{customlegend}[1][]{%
    \begingroup
    \csname pgfplots@init@cleared@structures\endcsname
    \pgfplotsset{#1}%
}{%
    \csname pgfplots@createlegend\endcsname
    \endgroup
}%

\def\addlegendimage{\csname pgfplots@addlegendimage\endcsname}

\pgfplotsset{
    jitter/.style={
        y filter/.code={\pgfmathparse{\pgfmathresult+rnd*#1}}
    },
    jitter/.default=0.05
}

\newcommand{\nop}[1]{}

\newcommand*{\eg}{{\em e.g.}}

\newcommand*{\ie}{{\em i.e.}}

\makeatletter
\def\@IEEEpubidpullup{8\baselineskip}
\makeatother

\begin{document}

\IEEEoverridecommandlockouts

\IEEEpubidadjcol

\title{\LARGE{\textbf{Library Adoption Dynamics in Software Teams}}}

\author{\IEEEauthorblockN{Pamela Bilo Thomas}
\IEEEauthorblockA{\textit{University of Notre Dame}
\texttt{pthomas4@nd.edu}\\
}
\and
\IEEEauthorblockN{Rachel Krohn}
\IEEEauthorblockA{\textit{University of Notre Dame}
\texttt{rkrohn@nd.edu}}\\
\and
\IEEEauthorblockN{Tim Weninger}
\IEEEauthorblockA{
\textit{University of Notre Dame}
\texttt{tweninge@nd.edu}
}
}

\maketitle
\begin{abstract}
When a group of people strives to understand new information, struggle ensues as various ideas compete for attention.  Steep learning curves are surmounted as teams learn together.  To understand how these team dynamics play out in software development, we explore Git logs, which provide a complete change history of software repositories.  In these repositories, we observe code additions, which represent successfully implemented ideas, and code deletions, which represent ideas that have failed or been superseded. By examining the patterns between these commit types, we can begin to understand how teams adopt new information. We specifically study what happens after a software library is adopted by a project, \ie, when a library is used for the first time in the project. We find that a variety of factors, including team size, library popularity, and prevalence on Stack Overflow are associated with how quickly teams learn and successfully adopt new software libraries.
\end{abstract}

\section{Introduction}

The process of learning new information and adopting new technology can be challenging. When new information is acquired, a learning curve often exists until an individual or group becomes proficient in the new technology. These challenges are present throughout society but are especially prevalent in computing and software development, where existing technologies are ever-changing and new innovations are constantly being developed. Further complicating these issues, most major software development occurs in teams where multiple parties collaborate on a project together, and where teammates may or may not know each other. This can lead to teammates having communication issues, which could result in teams which are inefficient and uncooperative. Online collaboration systems like GitHub have provided a powerful setting in which to study this process, because through analyzing the history of commits, whether removed, added, or otherwise, we can reconstruct a story of what happened between the teammates to eventually create the finalized code. By retelling this story, we can further investigate the struggles that occurred as the teammates tried to learn new information together, and see how long it takes for a team to become totally proficient at working together using the same technology.

The present work introduces a new approach to study this problem. By investigating how software developers adopt and use software libraries in this specific context, we may better understand how humans learn new technical information and incorporate concepts previously unknown to the user or group. The findings from this study can be generalized to understand how humans work, learn, and fight together, since GitHub provides a rich dataset which approximates the collaborative process.

Previous work by Kula et al found that programmers adopt new libraries only when they can be assured of the library's quality and functional correctness \cite{kula2015trusting}. But what happens after the adoption event? When are other group member receptive to new libraries; and when do they resist? What tools can help a team find and learn how to use new libraries? How long does it take a group of people to be successful at learning new information together? And finally, what does it even mean for that group to be successful?

To answer these and other questions, we explore the circumstances surrounding a library adoption, including the number of commits, the size of commits (measured by lines of code), and other related information, including availability of online resources, such as Stack Overflow. We present an in-depth look at commit addition and deletions, and analyze these questions from a variety of different angles.

Additionally, we ask if there exists any competition among team members over the inclusion of a library. When teammates have disagreements about what should be included in a GitHub project, there will ultimately be a winner. Uncovering which user eventually wins these code fights is an interesting research question. We explore competition by examining \textit{code fights} where two users revert each other’s code over the course of several commits. Like edit-wars on Wikipedia~\cite{viegas2004studying}, by looking at the users and libraries that participate in code fights we can learn a great deal about the adoption and diffusion of information. Although the present work focuses specifically on code contributed to public Python projects hosted on GitHub, we hope that other domains can use the methodology of the present work in other explorations of information adoption generally.

In addition to competition, we also aim to find the characteristics of fast adoption.  It is preferable to have a short adoption period - in that way, teams can become productive more quickly, since a longer adoption time results in periods of lost production.  By finding what combinations of repositories and libraries result in quick adoption times, we can begin to understand how team members work together to learn new information. Therefore, we hope to identify the qualities that create a good team, and what the characteristics of a bad team are. We hope that the findings from our research can be applied to help improve team dynamics and create groups that can work well together.

To summarize, this work aims to answer the following research questions:

\paragraph{What are the events that happen when a team adopts a library for the first time?}
Learning new information together can be challenging. As a team strives to use a new library together efficiently and correctly, it is one of the aims of this work to uncover the events that happen as the group learns, and what causes groups to work together well (or poorly!). In the data, we hope to find when library adoptions are successful - and what leads team members to abandon library adoptions as they decide to use other libraries instead.

\paragraph{Are commits containing new libraries more likely to have deletions than other types of commits?}
One of our research questions is to understand how teams learn to use these new libraries. We expect to find that as teams struggle to use new libraries, commits that contain new libraries will contain many deletions as users attempt to understand how the library works. Eventually, as team members become proficient in using a library, and the library becomes established in a project, we expect to see that there will be less deletions in the project repository code relative to when the library was first adopted. This ratio of positive to negative commits will help us define what it means for a library to be adopted.

\paragraph{Do the answers to these questions vary by library type, team size, or the amount of information available on Stack Overflow?}
Not all libraries and teams are alike, and we expect to find that as the types of the teams and the libraries change, so will the adoption time. Therefore, we will analyze library adoptions along the axes of team size and library type, and discover the relationship between adoption speeds and resources available on places like StackOverflow. We attempt to show how the speed of library adoptions changes with the addition of easy to use online resources, which we use as a proxy to represent library popularity. We hypothesize that if there exist few resources to learn how to use a library, the time to adoption for that library will be longer than for those libraries that are more well documented.

\paragraph{Do team members fight over the adoption and usage of a new library?}
We cannot assume that all teams work together well. Fights between team members might result as teammates have differing opinions about what libraries to use or how to implement new code. Therefore, we analyze code fights as part of this work. We wish to uncover what happens in teams when teammates remove code, or remove entire libraries. We examine which libraries are most often fought over, and which team members end up winning code fights. We hypothesize that more experienced team members will win these code fights.

To answer these questions we cloned 259,613 repositories from Github. This data provided us entire commit histories with userids, timestamps and code-diffs across all of the projects. We also downloaded, indexed, and cross-referenced library usage within Stack Overflow. Library adoption in software repositories reveals a natural experiment from which causal relationships between inferred. By following a careful extraction and experimental methodology, the following pages presents answers to these questions and an exploration of this rich resource.

\begin{figure}[t]
\footnotesize{
\begin{tabular}{l@{\hspace{0.15em}}l@{}l@{\hspace{0em}}l@{}l}
1 & \texttt{+} & \textcolor{green}{\texttt{import numpy as np}}\\
2 & \texttt{+} & \textcolor{green}{\texttt{from numpy import random as rnd}}\\
3 & \texttt{-} & \textcolor{red}{\texttt{import math}}\\
4 &  & \texttt{import pandas as pd}\\
5 &  & \texttt{def print\_groups(df):}\\
6 & \texttt{-} & \textcolor{red}{\texttt{\qquad print('function not implemented')}}\\
7 & \texttt{+} & \textcolor{green}{\texttt{\qquad bins = np.linspace(df.a.min(),df.a.max(), 10)}} \\
8 & \texttt{+} & \textcolor{green}{\texttt{\qquad groups = df.groupby(np.digitize(df.a, bins))}}\\
9 & \texttt{+} & \textcolor{green}{\texttt{\qquad print(groups.mean())}}\\ \\
\end{tabular}
}
%\vspace{0.5cm}
    \caption{An example of a Git history log where lines which begin with a `+' represent additions, and lines which begin with a `-' represent deletions.}
    \label{fig:code}
\end{figure}

\section{Data and Methodology}
\paragraph{Characteristics of GitHub.}

Since GitHub's rise to popularity and the growth of the open source software movement, developers have become increasingly comfortable contributing to public repositories~\cite{dias2016does}. Riding this wave of easily accessible data, researchers have done considerable investigation into GitHub, but many of these studies are limited to a few dozen project repositories in small-scale experiments, thereby limiting their potential generalizability~\cite{cosentino2016findings}. Instead, our goal is to gather as much data as possible, which, as we shall see, presents different challenges. We summarize our findings from this large-scale collection of data in the subsequent sections of this paper. As future work, further, more specific analysis can be done on smaller datasets, but for the purposes of our research, we wish to treat this work as a big data problem, and try to ingest as much data as possible to answer these questions.

\paragraph{Project Repositories on GitHub}

The confluence of GitHub's online social system and Git's complete history of pull and commit behavior provides an unprecedented source of information on adoption behavior~\cite{kalliamvakou2014promises,mcdonnell2013empirical}. We focus on commits that contain imported libraries and their functions. The example code snippet in Fig.~\ref{fig:code} shows how a single commit can add and/or remove one or more lines of code. In this example, the 5 additions and 2 deletions result in a net commit size of +3 lines, including the adoption of \texttt{numpy}. We consider usages of libraries $\ell$ (\eg, \texttt{numpy}, \texttt{pandas}, \texttt{math}) that are imported into Python code through an \texttt{import} statement. An important assumption that we make is that submodule \texttt{import} statements represent the parent library. Therefore, submodules like \texttt{numpy.random} are considered equivalent to \texttt{numpy} in the present work, because the function is included as a part of the parent \texttt{numpy} library, and we are interested in functions which refer to the adopted library.

We also consider direct function calls on the imported library. It is important to carefully consider the scoping of library aliases (\eg, \texttt{np}, \texttt{pd}, \texttt{rnd} in Fig.~\ref{fig:code}) so that we can mine these library functions accurately. Thankfully, the library aliases are included in the import statement, so we can easily find these aliases in the code. In the above example, the functions from the \texttt{numpy} library (\texttt{linspace} and \texttt{digitize}) are referenced in two of the added lines. 

Indirect library usage is exemplified by the \texttt{groupby} function of the \texttt{df} object and by the \texttt{mean} function of the \texttt{groups} object in Fig.~\ref{fig:code}. These objects were created from some ancestor call to the \texttt{pandas} library (\eg, \texttt{df = pd.DataFrame(x)}) in some other location, but do not directly reference the \texttt{pandas} library themselves. We do not consider indirect library usage in the present work, and only focus on direct library calls.

\paragraph{Data Collection}

First, we issued a query to the GitHub search API for projects written primarily in Python. GitHub returned repository IDs of the 1,000 most popular Python projects on the site. We then found all GitHub users who made at least one commit to a repository in this set and retrieved all their Python projects. We did this breadth-first style crawling two additional times, culminating in 259,923 projects with 89,311 contributing GitHub users.

Of these, we were able to clone 259,613 Python repositories to disk; the remainder were made private or deleted between the time we crawled their project URL and the time that we performed the clone operation. Each cloned repository includes all files, branches, and versions, along with a complete edit history. These repositories constitute about 13\% of all Python repositories on GitHub as of September 2018. The full dataset of cloned projects occupies about 8 TB of disk space. For analysis, we parsed the commits to only contain imported functions and libraries, which drastically reduced the size of the dataset.

Because we sampled 13\% of the total available public Python projects available on GitHub, it is important to be wary of sampling biases. Our initial GitHub query returned the most popular projects, so our dataset may over-represent highly active repositories compared to the population. It is not our intention to faithfully survey GitHub use nor to represent all Python projects; instead, our goal is to understand how programmers adopt and use new software libraries. Our findings can be applied to projects of all sizes, and small projects are well-represented in our data. However, it is important to remember that private software repositories are not included in our dataset, so we can only investigate team interactions in a public environment.

Additionally, we downloaded all question posts from Stack Overflow from its inception until September 2018. Appropriate tagging tends to increase viewership of questions~\cite{saha2013discriminative}, so we filtered out any posts that were not tagged as Python posts, then extracted all libraries from any code block using the same pattern matching technique as used in the library extraction from Python projects. Only top-level libraries were included. Because Stack Overflow is a free-form text entry platform, this pattern matching procedure may occasionally count extraneous words and misspellings as libraries. Additionally, it is possible that some libraries might not be returned by our query because the library name might have been misspelled, or the question did not include Python code containing a library import. 

\paragraph{Recreating the Project Workflow}
%\begin{figure}[t]
%    \centering
%        \include{./figs/commits_by_hour_month}
%    \caption{Total lines of code referencing newly-adopted library $\ell$ as a function of time. Adoption of $\ell$ occurs at $x=0$, meaning that time is measured as hours after the first commit. The cyclic (\ie, up and down) pattern corresponds to day-night development cycles, since we can assume that more commits happen during the day, and then work resumes at approximately the same time the next day after a period of rest. Most commits that reference $\ell$ occur within the first few days and then stabilize. For this image, we plot a count of all commits that occur across all projects and libraries.}
%    \label{fig:commitsbyhour}
%\end{figure}

Each project repository contains files, commits, and users. Each commit is marked with a timestamp $t$, a message, one (or many) parent-commits (in the event of a merge operation), and a \texttt{diff} specifying lines that were added, edited, or deleted within each file.

An important complication that arises in Git commit histories is that stashes, reverts, branches, and other Git commands can result in a non-monotonic chronology of edits. Because of this, we should not compare commits across different branches to each other until they are merged. Reversions introduce an additional complication. For example, if a user (1) commits code on Monday, (2) commits some new code on Wednesday, and then (3) on Friday reverts the project's code to Monday's commit, then the chronology of the repository flows non-monotonically from Monday to Wednesday to Monday again despite the reversion being made on Friday. 

Fortunately, each commit keeps a pointer to its parent(s), so we can create the actual lineage of the commits by following the graph of each commit's parent rather than blindly following the commit timestamps. Because the order of actions is more important than exact times, we enforce a monotonic chronology according to the commit graph. Future work can attempt to explore how this problem can be approached by using timestamps to analyze how the project changes over time.

\begin{figure*}[t]
    \centering
    \begin{subfigure}{0.25\textwidth}
        \include{./figs/commit_dist}
        \caption{\label{fig:commit_dist}}
    \end{subfigure}
    \begin{subfigure}{0.25\textwidth}
        \pgfplotstableread{
x   px
1	0.032710153
2	0.046165393
3	0.060562227
4	0.065938865
5	0.065215611
6	0.063905568
7	0.060930677
8	0.053752729
9	0.047857533
10	0.04470524
11	0.039737991
12	0.035848799
13	0.031782205
14	0.027497271
15	0.025423035
16	0.022134279
17	0.021219978
18	0.018736354
19	0.016007096
20	0.014901747
21	0.013509825
22	0.012172489
23	0.011203603
24	0.010480349
25	0.009634279
26	0.008201419
27	0.007218886
28	0.006700328
29	0.006686681
30	0.006345524
31	0.005594978
32	0.005076419
33	0.004271288
34	0.004257642
35	0.004107533
36	0.003588974
37	0.003397926
38	0.003056769
39	0.003316048
40	0.002442686
41	0.002906659
42	0.002415393
43	0.002101528
44	0.002374454
45	0.002046943
46	0.001883188
47	0.001582969
48	0.002033297
49	0.001869541
50	0.001419214
51	0.001378275
52	0.001664847
53	0.001514738
54	0.001405568
55	0.00143286
56	0.001296397
57	0.001337336
58	0.000927948
59	0.001118996
60	0.001050764
61	0.000914301
62	0.000887009
63	0.000805131
64	0.00095524
65	0.000614083
66	0.000559498
67	0.000436681
68	0.00058679
69	0.000518559
70	0.000436681
71	0.00058679
72	0.000559498
73	0.000409389
74	0.000463974
75	0.000641376
76	0.000559498
77	0.000450328
78	0.000545852
79	0.000504913
80	0.000450328
81	0.000300218
82	0.000436681
83	0.000354803
84	0.000409389
85	0.00036845
86	0.000286572
87	0.000245633
88	0.000272926
89	0.000300218
90	0.000259279
91	0.000272926
92	0.00047762
93	0.000204694
94	0.000231987
95	0.000204694
96	0.000245633
97	0.000259279
98	0.000259279
99	0.000272926
100	0.000313865
101	0.000272926
102	0.000245633
103	0.000245633
104	0.000245633
105	0.000327511
106	0.000231987
107	0.000150109
108	0.000191048
109	0.000231987
110	0.000259279
111	0.000150109
112	0.000231987
113	0.000218341
114	0.000191048
115	0.000191048
116	0.000177402
117	0.000259279
118	0.00010917
119	8.18777E-05
120	0.000218341
121	0.00010917
122	0.000204694
123	0.000163755
124	0.000191048
125	0.000204694
126	0.000150109
127	0.000300218
128	0.00010917
129	0.000286572
130	0.00010917
131	0.000204694
132	9.5524E-05
133	0.000150109
134	5.45852E-05
135	0.000150109
136	0.000163755
137	0.000163755
138	0.000136463
139	0.000245633
140	0.000122817
141	0.000150109
142	0.000191048
143	0.000177402
144	0.000122817
145	0.000163755
146	0.00010917
147	0.000204694
148	0.000177402
149	9.5524E-05
150	6.82314E-05
151	4.09389E-05
152	0.000150109
153	4.09389E-05
154	9.5524E-05
155	0.000122817
156	2.72926E-05
157	5.45852E-05
158	5.45852E-05
159	0.00010917
160	0.00010917
161	0.000122817
162	2.72926E-05
163	8.18777E-05
164	6.82314E-05
165	0.000122817
166	9.5524E-05
167	2.72926E-05
168	8.18777E-05
169	9.5524E-05
170	9.5524E-05
171	8.18777E-05
172	4.09389E-05
173	5.45852E-05
174	4.09389E-05
175	9.5524E-05
176	4.09389E-05
177	0.00010917
178	9.5524E-05
179	4.09389E-05
180	4.09389E-05
181	0.000122817
182	9.5524E-05
183	5.45852E-05
184	5.45852E-05
185	6.82314E-05
186	0.000122817
187	8.18777E-05
188	4.09389E-05
190	6.82314E-05
191	0.000122817
192	8.18777E-05
193	4.09389E-05
194	0.000122817
195	4.09389E-05
196	1.36463E-05
197	5.45852E-05
198	5.45852E-05
200	8.18777E-05
201	5.45852E-05
202	4.09389E-05
203	9.5524E-05
204	4.09389E-05
205	5.45852E-05
206	6.82314E-05
207	4.09389E-05
208	4.09389E-05
209	6.82314E-05
210	5.45852E-05
211	9.5524E-05
212	5.45852E-05
213	1.36463E-05
214	2.72926E-05
215	2.72926E-05
216	4.09389E-05
217	4.09389E-05
218	2.72926E-05
219	2.72926E-05
220	9.5524E-05
221	1.36463E-05
222	4.09389E-05
223	9.5524E-05
224	1.36463E-05
225	1.36463E-05
227	0.00010917
228	4.09389E-05
229	4.09389E-05
230	4.09389E-05
231	1.36463E-05
232	4.09389E-05
233	2.72926E-05
234	2.72926E-05
235	8.18777E-05
236	4.09389E-05
237	4.09389E-05
238	8.18777E-05
239	2.72926E-05
240	4.09389E-05
241	4.09389E-05
242	2.72926E-05
244	2.72926E-05
246	6.82314E-05
247	2.72926E-05
248	4.09389E-05
249	1.36463E-05
250	4.09389E-05
251	1.36463E-05
252	1.36463E-05
253	1.36463E-05
254	4.09389E-05
255	4.09389E-05
256	4.09389E-05
257	4.09389E-05
259	1.36463E-05
260	1.36463E-05
261	2.72926E-05
262	1.36463E-05
263	4.09389E-05
264	4.09389E-05
266	2.72926E-05
267	1.36463E-05
268	2.72926E-05
269	4.09389E-05
270	1.36463E-05
271	1.36463E-05
272	2.72926E-05
273	2.72926E-05
274	4.09389E-05
275	2.72926E-05
276	6.82314E-05
277	2.72926E-05
278	1.36463E-05
279	2.72926E-05
280	4.09389E-05
281	4.09389E-05
283	6.82314E-05
284	2.72926E-05
285	1.36463E-05
286	4.09389E-05
287	1.36463E-05
288	2.72926E-05
289	2.72926E-05
290	4.09389E-05
291	1.36463E-05
292	2.72926E-05
293	2.72926E-05
294	1.36463E-05
295	1.36463E-05
296	2.72926E-05
297	5.45852E-05
298	4.09389E-05
299	5.45852E-05
300	2.72926E-05
301	1.36463E-05
302	2.72926E-05
303	1.36463E-05
304	4.09389E-05
305	4.09389E-05
306	5.45852E-05
307	6.82314E-05
308	1.36463E-05
309	8.18777E-05
311	4.09389E-05
312	1.36463E-05
313	5.45852E-05
314	1.36463E-05
315	1.36463E-05
316	8.18777E-05
317	5.45852E-05
318	5.45852E-05
319	4.09389E-05
320	4.09389E-05
321	2.72926E-05
322	4.09389E-05
323	1.36463E-05
324	5.45852E-05
326	8.18777E-05
327	1.36463E-05
328	4.09389E-05
329	2.72926E-05
330	2.72926E-05
331	1.36463E-05
332	4.09389E-05
333	4.09389E-05
336	5.45852E-05
337	2.72926E-05
338	1.36463E-05
339	2.72926E-05
340	4.09389E-05
341	5.45852E-05
343	1.36463E-05
344	4.09389E-05
345	2.72926E-05
347	2.72926E-05
349	1.36463E-05
350	1.36463E-05
351	5.45852E-05
352	2.72926E-05
353	1.36463E-05
356	2.72926E-05
357	1.36463E-05
358	4.09389E-05
359	5.45852E-05
360	6.82314E-05
361	1.36463E-05
362	1.36463E-05
363	4.09389E-05
364	2.72926E-05
365	4.09389E-05
366	1.36463E-05
367	1.36463E-05
368	1.36463E-05
369	2.72926E-05
370	1.36463E-05
372	4.09389E-05
374	2.72926E-05
375	1.36463E-05
376	1.36463E-05
377	4.09389E-05
378	1.36463E-05
380	1.36463E-05
381	5.45852E-05
382	6.82314E-05
383	2.72926E-05
384	4.09389E-05
385	1.36463E-05
386	2.72926E-05
387	1.36463E-05
389	5.45852E-05
390	2.72926E-05
391	2.72926E-05
392	1.36463E-05
393	1.36463E-05
394	2.72926E-05
395	2.72926E-05
396	2.72926E-05
397	2.72926E-05
398	4.09389E-05
399	2.72926E-05
400	1.36463E-05
405	2.72926E-05
406	4.09389E-05
407	1.36463E-05
408	4.09389E-05
410	6.82314E-05
412	2.72926E-05
413	5.45852E-05
414	1.36463E-05
415	1.36463E-05
416	2.72926E-05
417	2.72926E-05
418	1.36463E-05
420	1.36463E-05
424	1.36463E-05
425	2.72926E-05
426	1.36463E-05
427	1.36463E-05
428	2.72926E-05
430	1.36463E-05
433	5.45852E-05
435	1.36463E-05
436	4.09389E-05
437	2.72926E-05
438	2.72926E-05
439	2.72926E-05
440	2.72926E-05
441	2.72926E-05
442	1.36463E-05
445	4.09389E-05
446	2.72926E-05
447	2.72926E-05
449	1.36463E-05
450	2.72926E-05
452	1.36463E-05
453	1.36463E-05
454	2.72926E-05
456	1.36463E-05
457	1.36463E-05
460	1.36463E-05
463	1.36463E-05
464	1.36463E-05
465	1.36463E-05
468	1.36463E-05
469	4.09389E-05
470	4.09389E-05
473	1.36463E-05
476	1.36463E-05
478	4.09389E-05
479	5.45852E-05
480	1.36463E-05
483	1.36463E-05
488	1.36463E-05
489	2.72926E-05
492	2.72926E-05
493	1.36463E-05
494	1.36463E-05
497	1.36463E-05
500	2.72926E-05
501	1.36463E-05
504	2.72926E-05
505	1.36463E-05
508	1.36463E-05
510	1.36463E-05
511	2.72926E-05
513	1.36463E-05
516	1.36463E-05
518	2.72926E-05
519	1.36463E-05
521	1.36463E-05
524	2.72926E-05
525	2.72926E-05
527	1.36463E-05
530	1.36463E-05
533	1.36463E-05
535	2.72926E-05
536	5.45852E-05
537	2.72926E-05
538	1.36463E-05
539	1.36463E-05
540	1.36463E-05
541	1.36463E-05
543	1.36463E-05
544	1.36463E-05
545	2.72926E-05
554	1.36463E-05
555	1.36463E-05
556	1.36463E-05
558	1.36463E-05
559	1.36463E-05
560	1.36463E-05
565	2.72926E-05
566	1.36463E-05
568	1.36463E-05
569	1.36463E-05
570	1.36463E-05
577	1.36463E-05
578	1.36463E-05
584	1.36463E-05
589	1.36463E-05
591	2.72926E-05
597	1.36463E-05
598	1.36463E-05
601	1.36463E-05
604	1.36463E-05
605	1.36463E-05
607	1.36463E-05
609	1.36463E-05
614	1.36463E-05
618	1.36463E-05
619	1.36463E-05
624	1.36463E-05
626	5.45852E-05
628	1.36463E-05
629	1.36463E-05
630	1.36463E-05
633	1.36463E-05
637	1.36463E-05
641	1.36463E-05
647	1.36463E-05
649	1.36463E-05
657	1.36463E-05
658	1.36463E-05
662	2.72926E-05
663	1.36463E-05
664	8.18777E-05
666	1.36463E-05
668	1.36463E-05
670	5.45852E-05
680	1.36463E-05
685	1.36463E-05
695	1.36463E-05
699	1.36463E-05
701	2.72926E-05
728	1.36463E-05
759	1.36463E-05
774	1.36463E-05
782	1.36463E-05
785	1.36463E-05
796	1.36463E-05
812	1.36463E-05
818	1.36463E-05
820	1.36463E-05
828	1.36463E-05
836	1.36463E-05
837	1.36463E-05
855	1.36463E-05
856	1.36463E-05
859	1.36463E-05
878	1.36463E-05
895	1.36463E-05
906	1.36463E-05
909	2.72926E-05
942	1.36463E-05
956	1.36463E-05
995	1.36463E-05
1103	1.36463E-05
1170	1.36463E-05
1193	1.36463E-05
1349	1.36463E-05
1474	1.36463E-05
}{\adoptions}
\begin{tikzpicture}
    \begin{axis}[
    title=Adoption Distribution,
    clip=false,
    width=1.90in,
    height=1.75in,
    ymode=log,
    xmode=log,
    xlabel = {Libraries},
    ylabel = {$p(x)$},
    y label style={at={(axis description cs:-0.2,.5)}},
    ]
    \addplot [only marks, mark=*, blue, mark size=0.75pt, fill opacity=0.5, draw opacity=0.2]  table [x=x, y=px]   {\adoptions};
    \node[anchor=north east] at (rel axis cs:-0.05,-0.01) {\textbf{(b)}};
    \end{axis}
\end{tikzpicture}
        \caption{\label{fig:adopt_dist}}
    \end{subfigure}
    \begin{subfigure}{0.48\textwidth}
        \pgfplotstableread{
x  mean  median  sum  stdev  q1  q3  cnt  inf sup
0	6.605275739	2	1713250	29.53497286	0	6	259376	6.491610478	6.718941
1	3.318329405	0	805332	17.19877578	0	3	242692	3.249902661	3.386756149
2	1.136170136	0	251414	9.233688399	0	1	221282	1.097696942	1.17464333
3	0.821645815	0	166684	10.19798266	0	0	202866	0.777268022	0.866023609
4	0.658405966	0	123434	7.325564945	0	0	187474	0.625245038	0.691566893
5	0.558639336	0	97320	6.759705349	0	0	174209	0.526896294	0.590382377
6	0.491375231	0	79961	5.527671934	0	0	162729	0.464517715	0.518232748
7	0.471273165	0	72044	5.3867295	0	0	152871	0.44426975	0.49827658
8	0.418197568	0	60348	5.037670789	0	0	144305	0.39220524	0.444189895
9	0.420811034	0	57479	5.507455476	0	0	136591	0.391603424	0.450018645
10	0.384562594	0	49872	4.745755664	0	0	129685	0.35873306	0.410392128
11	0.337856767	0	41727	3.767128541	0	0	123505	0.316846855	0.358866679
12	0.360991946	0	42579	4.902560797	0	0	117950	0.333013081	0.38897081
13	0.330652447	0	37330	4.395977501	0	0	112898	0.305009472	0.356295422
14	0.319436093	0	34532	4.195063877	0	0	108103	0.294428277	0.34444391
15	0.312781651	0	32483	3.602763282	0	0	103852	0.290869532	0.334693769
16	0.289635153	0	28936	3.196735731	0	0	99905	0.269812162	0.309458145
17	0.272603607	0	26252	2.629678632	0	0	96301	0.255994613	0.289212602
18	0.290545337	0	27012	3.590406153	0	0	92970	0.267465739	0.313624935
19	0.273420965	0	24545	3.363360898	0	0	89770	0.251418875	0.295423054
20	0.271911822	0	23584	3.536072314	0	0	86734	0.248378539	0.295445106
21	0.263552136	0	22131	3.37352611	0	0	83972	0.240734382	0.286369891
22	0.241249585	0	19623	3.110640375	0	0	81339	0.219872107	0.262627063
23	0.267887677	0	21150	4.600731017	0	0	78951	0.235795137	0.299980218
24	0.250117481	0	19161	3.763756616	0	0	76608	0.223464822	0.27677014
25	0.251926226	0	18768	3.602672027	0	0	74498	0.226055508	0.277796945
26	0.235817805	0	17085	4.010727597	0	0	72450	0.206612615	0.265022996
27	0.244785981	0	17265	3.812491939	0	0	70531	0.216649162	0.272922799
28	0.232965465	0	16001	3.508044865	0	0	68684	0.206729719	0.259201211
29	0.214549312	0	14360	3.008673203	0	0	66931	0.191755475	0.237343149
30	0.226198877	0	14745	3.009209974	0	0	65186	0.203097844	0.24929991
31	0.236097791	0	15017	4.24811332	0	0	63605	0.203083153	0.269112429
32	0.21187313	0	13173	2.69106621	0	0	62174	0.190719931	0.233026329
33	0.242957283	0	14765	4.611932631	0	0	60772	0.20628928	0.279625286
34	0.20688785	0	12297	3.023498101	0	0	59438	0.18258072	0.231194979
35	0.20466397	0	11892	3.231141325	0	0	58105	0.178391236	0.230936705
36	0.192855886	0	10960	2.763428174	0	0	56830	0.170135515	0.215576257
37	0.208254048	0	11601	3.338987236	0	0	55706	0.180525965	0.235982131
38	0.199105719	0	10865	3.05576134	0	0	54569	0.173466635	0.224744803
39	0.19606669	0	10478	3.139763793	0	0	53441	0.169446218	0.222687163
40	0.203624835	0	10662	2.833685436	0	0	52361	0.179352937	0.227896734
41	0.18609634	0	9554	2.793224026	0	0	51339	0.161934047	0.210258633
42	0.167129924	0	8418	1.847318429	0	0	50368	0.150996725	0.183263122
43	0.17015728	0	8417	2.106908858	0	0	49466	0.151589994	0.188724566
44	0.192287092	0	9334	2.618275449	0	0	48542	0.168994781	0.215579402
45	0.201623561	0	9612	4.26298649	0	0	47673	0.163355735	0.239891386
46	0.159187943	0	7457	1.264518921	0	0	46844	0.147736653	0.170639233
47	0.167958656	0	7735	2.210640437	0	0	46053	0.147768243	0.18814907
48	0.182268406	0	8244	3.245171516	0	0	45230	0.152360884	0.212175928
49	0.18701363	0	8315	6.030886983	0	0	44462	0.130954963	0.243072296
50	0.151960336	0	6651	1.048847099	0	0	43768	0.142134039	0.161786634
51	0.1681987	0	7246	2.12316463	0	0	43080	0.148149276	0.188248124
52	0.158836161	0	6731	2.816898036	0	0	42377	0.132015956	0.185656366
53	0.159097449	0	6635	2.370112255	0	0	41704	0.13634982	0.181845077
54	0.160984525	0	6606	1.926570488	0	0	41035	0.142343759	0.179625291
55	0.161014012	0	6504	2.697277766	0	0	40394	0.13470992	0.187318104
56	0.151122495	0	6018	2.35801967	0	0	39822	0.127962313	0.174282677
57	0.141142595	0	5539	1.145571203	0	0	39244	0.129808378	0.152476812
58	0.186837067	0	7222	3.060786449	0	0	38654	0.156323578	0.217350557
59	0.159364746	0	6071	2.711258975	0	0	38095	0.132138166	0.186591326
60	0.181585746	0	6818	3.638108901	0	0	37547	0.144786057	0.218385435
61	0.141274687	0	5229	0.95403046	0	0	37013	0.131555249	0.150994126
62	0.174931507	0	6385	3.60975111	0	0	36500	0.137898677	0.211964337
63	0.142841257	0	5138	2.190484652	0	0	35970	0.120203882	0.165478632
64	0.176097437	0	6246	4.298794341	0	0	35469	0.131359262	0.220835613
65	0.152913939	0	5350	1.796471665	0	0	34987	0.134089455	0.171738424
66	0.147344953	0	5089	1.880400798	0	0	34538	0.127513346	0.16717656
67	0.135570825	0	4617	1.086588722	0	0	34056	0.124030329	0.14711132
68	0.134609099	0	4530	1.165402013	0	0	33653	0.122157648	0.147060549
69	0.153922453	0	5121	3.474847416	0	0	33270	0.116583218	0.191261687
70	0.152974504	0	5022	2.611232914	0	0	32829	0.12472747	0.181221538
71	0.158321514	0	5135	3.359133351	0	0	32434	0.121763461	0.194879568
72	0.131320378	0	4210	1.214086268	0	0	32059	0.118030205	0.144610551
73	0.137053755	0	4342	2.148747987	0	0	31681	0.113392263	0.160715246
74	0.120246693	0	3763	1.185009365	0	0	31294	0.107117219	0.133376166
75	0.132645804	0	4103	1.920698898	0	0	30932	0.111240995	0.154050613
76	0.17775597	0	5434	4.830208596	0	0	30570	0.12360899	0.23190295
77	0.124871684	0	3771	0.916568409	0	0	30199	0.114533967	0.135209402
78	0.120726567	0	3609	0.913157281	0	0	29894	0.110374916	0.131078219
79	0.149690471	0	4425	2.864428937	0	0	29561	0.117036608	0.182344333
80	0.145289793	0	4249	2.894979847	0	0	29245	0.112109837	0.178469749
81	0.196398327	0	5682	6.243534043	0	0	28931	0.124452629	0.268344025
82	0.123846477	0	3543	1.446233101	0	0	28608	0.107087379	0.140605574
83	0.132996037	0	3759	1.181412498	0	0	28264	0.119222648	0.146769427
84	0.106278637	0	2969	0.698280153	0	0	27936	0.09809015	0.114467124
85	0.151872625	0	4197	1.476694172	0	0	27635	0.134461899	0.169283352
86	0.150498375	0	4122	3.868364356	0	0	27389	0.104684676	0.196312074
87	0.136697553	0	3703	3.718076289	0	0	27089	0.092420584	0.180974521
88	0.127797672	0	3426	1.102041004	0	0	26808	0.114605339	0.140990006
89	0.158785184	0	4214	4.083609763	0	0	26539	0.10965391	0.207916458
90	0.124515246	0	3275	2.383257929	0	0	26302	0.095712577	0.153317915
91	0.158655139	0	4129	4.628425363	0	0	26025	0.102421788	0.214888491
92	0.131175535	0	3380	3.228125629	0	0	25767	0.091759351	0.170591718
93	0.124700898	0	3179	0.946009998	0	0	25493	0.113087981	0.136313815
94	0.149787942	0	3779	4.662236819	0	0	25229	0.092257146	0.207318738
95	0.120241764	0	3004	1.175964335	0	0	24983	0.105659401	0.134824128
96	0.145133889	0	3588	2.277429765	0	0	24722	0.11674429	0.173523488
97	0.140866936	0	3448	1.804044731	0	0	24477	0.118266112	0.163467761
98	0.154223523	0	3741	4.795380423	0	0	24257	0.093875841	0.214571205
99	0.126023866	0	3031	1.398619241	0	0	24051	0.108347662	0.14370007
100	0.131880589	0	3141	2.833949581	0	0	23817	0.095888699	0.16787248
}{\table}
\begin{tikzpicture}
 \begin{groupplot}[
    group style={%
        group size=1 by 2,%
        x descriptions at=edge bottom,%
        vertical sep=0pt,%
    },
    clip=true,
    clip mode=individual,
    width=3.4in,
    xmin=0,
    xmax=100,
    ymode=log,
    legend cell align=left,
    legend pos=outer north east,
    legend style={draw=none}
    ]
    \nextgroupplot[
        title=Library Adoptions per commit,
        xticklabels={,,}, 
        height=1.5in, 
        ymin=0.05,
        y label style={at={(axis description cs:-0.08,.5)}},
        ylabel = {avg adopts},
        ]
    \addplot [stack plots=y, fill=none, draw=none, forget plot]   table [x=x, y=inf]   {\table} \closedcycle;
    \addplot [stack plots=y, fill=gray!70, opacity=0.6, draw opacity=1, thin, smooth, area legend]   table [x=x, y expr=\thisrow{sup}-\thisrow{inf}]   {\table} \closedcycle;
    \addplot [stack plots=false, blue, thick, smooth]  table [x=x, y=mean]   {\table};
    \nextgroupplot[
        ybar, 
        bar width=0.8pt, 
        ymin=100,
        ymode=log,
        height=0.85in, 
        y label style={at={(axis description cs:-0.1,.5)}},
        xlabel = {Commit \#},
        ylabel = {vol},
    ]
    \addplot []  table [x=x, y=sum]   {\table};
    \node[anchor=north east] at (rel axis cs:-0.05,-0.05) {\textbf{(c)}};
    \end{groupplot}
\end{tikzpicture}
        \caption{\label{fig:adopt_per_commit}}
    \end{subfigure}
    \vspace{-2\baselineskip}
    \caption{(a) Number of commits per project follows a shifted power law distribution, \ie, most projects have few commits, and few projects have many commits. (b) Number of libraries used per project also follows a shifted power law distribution, \ie, most projects adopt few libraries, and few libraries are adopted by many repositories. (c) When libraries are adopted into repositories, it tends to occur early in the repository history.}
    \label{fig:dists}
\end{figure*}
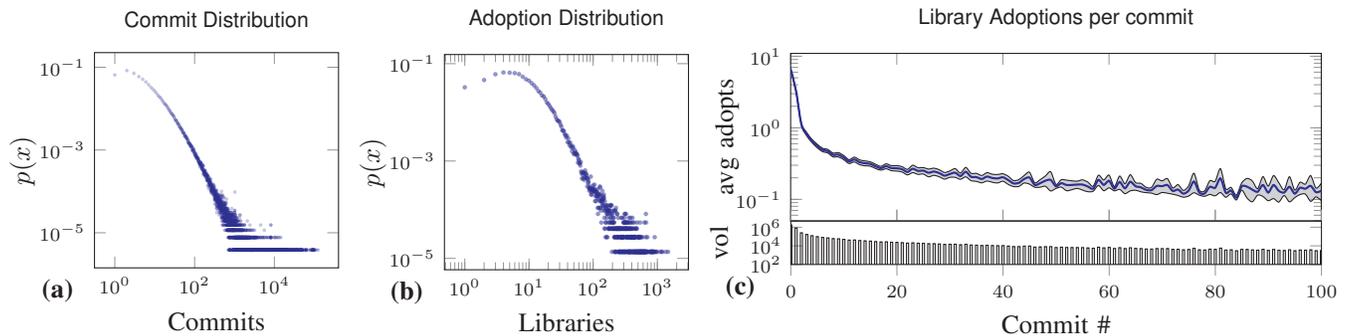

\paragraph{Text Mining for Libraries}

After downloading the data, we combed through each Git commit log to find which libraries had been imported by using Regex pattern matching to identify import statements, such as `from $\ell$ import $f$" and `import $\ell$ as $f$". Once we identified which libraries were used in a Git project, we searched the log to find the lines which referenced the functions contained in a library import. To do this, we used pattern matching to search for the libraries $\ell$ and functions $f$, along with indicators such as \texttt{.} and \texttt{(}, to indicate that the library or function was being used. We gathered the author name, and then stored all this information, including library used, author name, and commit type, in a database for further analysis. This parsing of the data allowed us to perform a relatively quick analysis on a smaller dataset than the entire commit log.

\section{Library Adoption}
Formally, for a project $p$, a library $\ell$, and a time $t$, we define a \textit{library adoption} to be an event ($p$, $\ell$, $t$) representing the first time $t$ that $\ell$ is found in $p$. 

Some project repositories are simple, containing only a single commit, while others are extremely complex with multiple branches, hundreds of users, and thousands of commits. In 2014, Kalliamvakou et al found that the median number of commits per project on a small sample of GitHub projects was 6, and 90 percent of projects had less than 50 commits~\cite{kalliamvakou2014promises}. Our dataset shows a slightly larger distribution of commits, though as mentioned earlier, it is possible that our GitHub data is over-represented by active projects. The distribution of commit-activity per project, illustrated in Fig.~\ref{fig:commit_dist}, resembles a shifted power law distribution. Because of this dynamic, 50\% of projects were found to have 10 or fewer commits (\ie, median of 10) and 90\% of projects have 100 or fewer commits.

The distribution of the number of libraries adopted per project, illustrated in Fig.~\ref{fig:adopt_dist}, also resembles a shifted power law distribution, albeit with a larger offset than the commit-activity distribution of Fig.~\ref{fig:commit_dist}. However, the number of adoptions is less evenly distributed: 54\% of projects adopted 10 or fewer distinct libraries and 98\% of projects adopted 100 or fewer libraries.

Across all commits of all projects, we find that library adoptions occur more frequently within the first few commits. Figure~\ref{fig:adopt_per_commit} shows that a project's first commit adopts 6.4 libraries on average (with a median of 2 not illustrated). A project's second through fifth commits adopt 3.3, 1.1, 0.8, and 0.65 libraries on average (with median values of 0 not illustrated). In general, the average number of adoptions per commit appears to follow a Zipfian decay, and commits tend to occur early in a project repository  history.

%we can see many different storylines begin to emerge regarding how software developers write and commit software and work together in teams. The goal of the current work is to examine events that surround the library adoptions for indications of changes in productivity or other behavior. 

\section{Activity Change after Adoption}

A simple (albeit poor) indicator of productivity in software projects is the number of lines of code (LOC) that are added and/or removed in a commit. While it can be argued that not every line added to a repository is important, productive, or useful, and that inefficient code tends to have more lines, we go with this indicator because it is the simplest to understand, and provides a good summary statistic. Further analysis can be done on measuring the effectiveness of lines of code added to a repository. Within a single project the addition of code lines typically indicates positive activity like the addition of new functionality or features. Conversely, the removal of code typically indicates some negative activity like reversions or mistakes that are removed by the user. Oftentimes software bugs or other types of mistakes occur requiring edits that both add and remove code. Our data will contain the number of lines that are added or deleted in each commit, along with the libraries involved.

We begin our analysis of library adoptions by parsing each code commit over each project repository. For each project starting from the first commit, we retrieve all imported libraries, their commit number, and the scope of any aliases. Users reference libraries and use them in several different ways. We define two classes of library use for the purposes of this study: (1) direct use, and (2) indirect use.

Direct library use, as the name implies, references the library name or alias directly. From the example in Fig. \ref{fig:code} above, the function call \texttt{np.linspace} directly references the \texttt{np} alias of the \texttt{numpy} library; this is an example of direct use. Indirect library use references library calls made through a variable or other intermediary. Again from the example in Fig.~\ref{fig:code} above, the function call \texttt{df.groupby} is an indirect use of the \texttt{pandas} library because it references the library's functions through the \texttt{df} variable. Taking an accurate accounting of indirect library use, especially in a language as fluid and dynamic as Python, would require an interpretation or partial-compile of each of the 23 million commits in our dataset. Therefore, in the present work, we limit our analysis to the import and direct use of libraries.

%
%\begin{figure}[t]
%    \centering
%        \include{./figs/afteradoption}
%    \caption{Mean average lines of code referencing $\ell$ grouped by additions, deletions and their net (additions-deletions) per commit after the adoption of $\ell$, where time 0 refers to when the library $\ell$ was referenced for the first time.}
%    \label{fig:afteradoption}
%\end{figure}
%

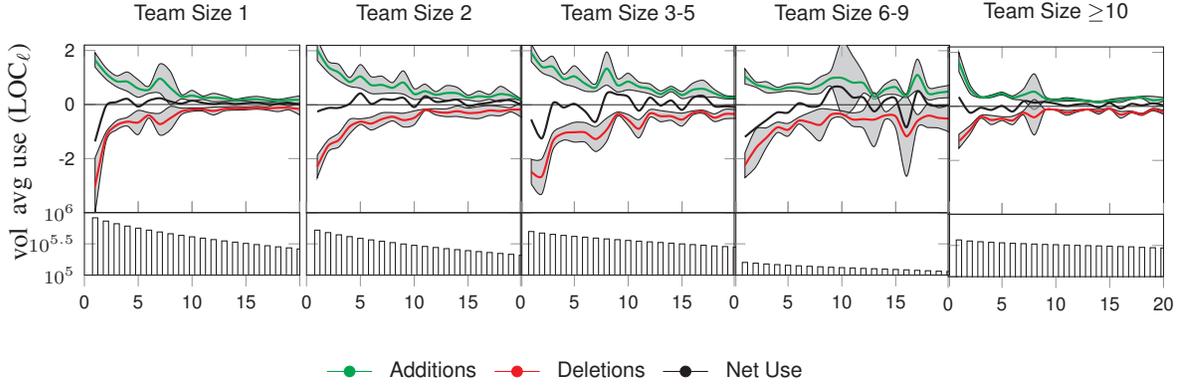
\begin{figure*}[t]
    \centering
    \begin{minipage}{0.205\textwidth}
        \pgfplotstableread{
x  mean  ci std len
1	1.651243319	0.268344826	124.3277306	824607
2	1.153490169	0.210896099	92.00373456	731091
3	0.859202326	0.203605688	84.67123565	664339
4	0.845469499	0.357550317	142.2404817	607953
5	0.592620728	0.324996396	124.4607879	563386
6	0.545660992	0.133491075	49.28309535	523588
7	0.958971715	0.53884848	192.418113	489843
8	0.612923462	0.604793682	209.0440725	458944
9	0.29140521	0.086978782	29.19234654	432723
10	0.329555317	0.193305189	63.14599522	409924
11	0.24744261	0.068060012	21.67601108	389649
12	0.257003862	0.15565089	48.30966897	370053
13	0.22087338	0.100502264	30.43465659	352277
14	0.160015077	0.042064247	12.44785979	336405
15	0.192968426	0.067738068	19.59923281	321597
16	0.193829651	0.107885226	30.54580517	307949
17	0.156632105	0.082862458	22.96657436	295105
18	0.151560542	0.038463885	10.4474618	283410
19	0.168961931	0.081880358	21.81353701	272642
20	0.170350674	0.149971232	39.17684697	262146

}{\pos}

\pgfplotstableread{
x  mean  ci std len
%0	0	0	0	0.0	0	0	0	-0.0	0.0	0.0
1	-3.019223195	1.037067792	480.4873152	824607
2	-1.220897439	0.185654728	80.99214888	731091
3	-0.768173507	0.171583405	71.35448453	664339
4	-0.636285262	0.163460274	65.02768156	607953
5	-0.666687904	0.447195715	171.2583023	563386
6	-0.399717351	0.100645653	37.15701069	523588
7	-0.725532819	0.402519277	143.7361383	489843
8	-0.502135465	0.432424618	149.4655218	458944
9	-0.258146222	0.124241371	41.69864257	432723
10	-0.189622433	0.064941517	21.21410588	409924
11	-0.18394583	    0.060704007	19.33324281	389649
12	-0.23826764	    0.146337754	45.41913276	370053
13	-0.161422978	0.072489645	21.95171907	352277
14	-0.138586022	0.067127551	19.86471635	336405
15	-0.142471116	0.056043959	16.21567653	321597
16	-0.120249894	0.051523314	14.58792056	307949
17	-0.143031603	0.064510151	17.87995693	295105
18	-0.120478455	0.141609999	38.46374422	283410
19	-0.123048765	0.075244055	20.04557661	272642
20	-0.169176235	0.215951944	56.41292785	262146

}{\neg}

\pgfplotstableread{
x  mean
1	-1.367979876
2	-0.06740727
3	0.091028819
4	0.209184237
5	-0.074067176
6	0.145943641
7	0.233438896
8	0.110787997
9	0.033258988
10	0.139932884
11	0.063496781
12	0.018736222
13	0.059450402
14	0.021429055
15	0.05049731
16	0.073579758
17	0.013600502
18	0.031082088
19	0.045913166
20	0.001174439

}{\mid}

\begin{tikzpicture}

\pgfplotsset{
every axis legend/.append style={
at={(0.5,-0.650)},
anchor=north
},
}
 
 \begin{groupplot}[
    group style={%
        group size=1 by 2,%
        x descriptions at=edge bottom,%
        vertical sep=0pt,%
    },
    clip=true,
    clip mode=individual,
    width=1.75in,
    xmin=0,
    xmax=20,
    legend columns=3,
    legend style={draw=none}
    %ymode=log,
    ]
    \nextgroupplot[
        title = Team Size 1,
        xticklabels={,,}, 
        height=1.5in, 
        ymax=2.2,
        ymin=-4,
        y label style={at={(axis description cs:-0.2,.5)}},
        ylabel = {avg use (LOC$_\ell$)},
        yticklabels={,,-2,0,2}
        ]
    \addplot [stack plots=y, fill=none, draw=none, forget plot]   table [x=x, y expr=(\thisrow{mean} + \thisrow{ci})]   {\pos} \closedcycle;
    \addplot [stack plots=y, fill=gray!70, opacity=0.6, draw opacity=1, thin, smooth, area legend]   table [x=x, y expr=(\thisrow{mean} - \thisrow{ci}) - (\thisrow{mean} + \thisrow{ci}) ]   {\pos} \closedcycle;
    \addplot [stack plots=y, stack dir=minus, forget plot, draw=none] table [x=x, y expr = (\thisrow{mean} - \thisrow{ci})] {\pos};

    \addplot [stack plots=y, fill=none, draw=none, forget plot]   table [x=x, y expr=(\thisrow{mean} + \thisrow{ci})]   {\neg} \closedcycle;
    \addplot [stack plots=y, fill=gray!70, opacity=0.6, draw opacity=1, thin, smooth, area legend]   table [x=x,y expr=(\thisrow{mean} - \thisrow{ci}) - (\thisrow{mean} + \thisrow{ci}) ]   {\neg} \closedcycle;
    \addplot [stack plots=y, stack dir=minus, forget plot, draw=none] table [x=x, y expr = (\thisrow{mean} - \thisrow{ci})] {\neg};
    
    \addplot [stack plots=false, green, thick, smooth]  table [x=x, y=mean]
    {\pos};
    \addplot [stack plots=false, red, thick, smooth]  table [x=x, y=mean]   
    {\neg};
    
    \draw[ultra thin] (axis cs:\pgfkeysvalueof{/pgfplots/xmin},0) -- (axis cs:\pgfkeysvalueof{/pgfplots/xmax},0);
    
    \addplot [stack plots=false, black, thick, smooth]  table [x=x, y=mean]   
    {\mid};\legend{}
    
    \nextgroupplot[
        ybar, 
        bar width=2pt, 
        ymode=log,
        height=0.95in, 
        y label style={at={(axis description cs:-0.22,.5)}},
        ylabel = {vol},
        ymax=1000000,
        ymin=100000,
        xticklabels={,0,5,10,15},
    ]
    \addplot []  table [x=x, y=len]   {\pos};
    %\node[anchor=north east] at (rel axis cs:-0.05,-0.05) {\textbf{(c)}};
    \end{groupplot}
\end{tikzpicture}
    \end{minipage}
    \begin{minipage}{0.15\textwidth}
        \pgfplotstableread{
x  mean  ci std len
1	2.032319155	0.390489148	144.0994078	523124
2	1.405734328	0.211513512	74.65391497	478550
3	1.19091508	0.303706364	102.91952	441150
4	0.843524292	0.164701502	54.01549606	413181
5	1.049010875	0.438143063	139.1126715	387258
6	0.729374049	0.313571174	96.78039564	365933
7	0.735853468	0.321403713	96.76260802	348187
8	0.627459823	0.194976046	57.23694008	331047
9	0.822568942	0.430662961	123.2008865	314378
10	0.396675389	0.199320836	55.68490723	299826
11	0.501847419	0.261122479	71.44219734	287555
12	0.370474706	0.116694921	31.30837487	276514
13	0.433413131	0.275542234	72.36399275	264953
14	0.424682663	0.21634018	55.77621183	255343
15	0.263675166	0.09076918	22.98185734	246260
16	0.31461233	0.192380559	47.87100008	237861
17	0.289943585	0.094040461	22.99555631	229699
18	0.396114018	0.286053622	68.77245987	222042
19	0.339614186	0.115063511	27.2254449	215068
20	0.161675574	0.029095274	6.777355727	208438

}{\pos}

\pgfplotstableread{
x  mean  ci std len
%0	0	0	0	0.0	0	0	0	-0.0	0.0	0.0
1	-2.28571243	0.41616334	153.5737704	523124
2	-1.52653101	0.217528745	76.77699748	478550
3	-1.30500965	0.259847234	88.05660908	441150
4	-0.808520766	0.268092976	87.92375827	413181
5	-0.626034592	0.104346962	33.13069606	387258
6	-0.696400735	0.237177896	73.20242586	365933
7	-0.531818097	0.23377678	70.38142363	348187
8	-0.444286085	0.154829675	45.45161833	331047
9	-0.56727072	0.253591731	72.54565383	314378
10	-0.48255391	0.257368661	71.90191606	299826
11	-0.188362469	0.028945017	7.919255562	287555
12	-0.261334113	0.113762551	30.52164181	276514
13	-0.295431095	0.14529487	38.15791415	264953
14	-0.239308911	0.19490955	50.25102761	255343
15	-0.307661545	0.185976598	47.08743263	246260
16	-0.288803214	0.224145662	55.77526697	237861
17	-0.21094069	0.183189545	44.79503245	229699
18	-0.234517314	0.173795418	41.78355911	222042
19	-0.181266934	0.084952093	20.1007123	215068
20	-0.180976506	0.293452808	68.35591561	208438

}{\neg}

\pgfplotstableread{
x  mean
1	-0.253393275
2	-0.120796682
3	-0.11409457
4	0.035003526
5	0.422976283
6	0.032973314
7	0.204035371
8	0.183173737
9	0.255298222
10	-0.085878521
11	0.31348495
12	0.109140593
13	0.137982036
14	0.185373752
15	-0.043986379
16	0.025809116
17	0.079002895
18	0.161596704
19	0.158347252
20	-0.019300932

}{\mid}

\begin{tikzpicture}

\pgfplotsset{
every axis legend/.append style={
at={(0.5,-0.650)},
anchor=north
},
}
 
 \begin{groupplot}[
    group style={%
        group size=1 by 2,%
        x descriptions at=edge bottom,%
        vertical sep=0pt,%
    },
    clip=true,
    clip mode=individual,
    width=1.75in,
    xmin=0,
    xmax=20,
    legend columns=3,
    legend style={draw=none}
    %ymode=log,
    ]
    \nextgroupplot[
        title = Team Size 2,
        xticklabels={,,}, 
        height=1.5in, 
        ymax=2.2,
        ymin=-4,
        y label style={at={(axis description cs:0.08,.5)}},
        yticklabels={,,},
        ]
    \addplot [stack plots=y, fill=none, draw=none, forget plot]   table [x=x, y expr=(\thisrow{mean} + \thisrow{ci})]   {\pos} \closedcycle;
    \addplot [stack plots=y, fill=gray!70, opacity=0.6, draw opacity=1, thin, smooth, area legend]   table [x=x, y expr=(\thisrow{mean} - \thisrow{ci}) - (\thisrow{mean} + \thisrow{ci}) ]   {\pos} \closedcycle;
    \addplot [stack plots=y, stack dir=minus, forget plot, draw=none] table [x=x, y expr = (\thisrow{mean} - \thisrow{ci})] {\pos};

    \addplot [stack plots=y, fill=none, draw=none, forget plot]   table [x=x, y expr=(\thisrow{mean} + \thisrow{ci})]   {\neg} \closedcycle;
    \addplot [stack plots=y, fill=gray!70, opacity=0.6, draw opacity=1, thin, smooth, area legend]   table [x=x,y expr=(\thisrow{mean} - \thisrow{ci}) - (\thisrow{mean} + \thisrow{ci}) ]   {\neg} \closedcycle;
    \addplot [stack plots=y, stack dir=minus, forget plot, draw=none] table [x=x, y expr = (\thisrow{mean} - \thisrow{ci})] {\neg};
    
    \addplot [stack plots=false, green, thick, smooth]  table [x=x, y=mean]
    {\pos};
    \addplot [stack plots=false, red, thick, smooth]  table [x=x, y=mean]   
    {\neg};
    
    \draw[ultra thin] (axis cs:\pgfkeysvalueof{/pgfplots/xmin},0) -- (axis cs:\pgfkeysvalueof{/pgfplots/xmax},0);
    
    \addplot [stack plots=false, black, thick, smooth]  table [x=x, y=mean]   
    {\mid};\legend{}
    
    \nextgroupplot[
        ybar, 
        bar width=2pt, 
        ymode=log,
        height=0.95in, 
        ymax=1000000,
        ymin=100000,
        yticklabels={,,},
        xticklabels={,0,5,10,15},
    ]
    \addplot []  table [x=x, y=len]   {\pos};
    %\node[anchor=north east] at (rel axis cs:-0.05,-0.05) {\textbf{(c)}};
    \end{groupplot}
\end{tikzpicture}
    \end{minipage}
    \begin{minipage}{0.15\textwidth}
        \pgfplotstableread{
x  mean  ci std len
1	1.929151	0.332327057	119.6493369	497952
2	1.402172379	0.216374955	75.75029227	470818
3	1.407359181	0.38379334	131.188097	448841
4	0.956451968	0.18120385	60.76246089	431951
5	1.052755592	0.389524024	128.2198413	416237
6	0.781025128	0.277145313	89.87104709	403946
7	0.641322247	0.140513969	44.82142703	390869
8	1.339085144	0.616543453	193.7219026	379253
9	0.724237334	0.251009178	77.81794458	369215
10	0.859390478	0.307410666	93.92918508	358643
11	0.667341455	0.174888026	52.67807978	348529
12	0.589870031	0.182939249	54.44278929	340225
13	0.393521774	0.119062476	35.02846136	332500
14	0.582539333	0.1006909	29.25498801	324279
15	0.360437629	0.132081504	37.86210685	315664
16	0.553248332	0.2436437	68.96718771	307804
17	0.58736119	0.314332604	87.91637578	300511
18	0.426870409	0.143010747	39.46750026	292578
19	0.283243411	0.064960025	17.7278448	286101
20	0.261339057	0.048572007	13.10275791	279546

}{\pos}

\pgfplotstableread{
x  mean  ci std len
%0	0	0	0	0.0	0	0	0	-0.0	0.0	0.0
1	-2.473264252	0.450608028	162.2346134	497952
2	-2.642127161	0.651327469	228.0219819	470818
3	-1.357298082	0.319545701	109.2269931	448841
4	-1.056686067	0.253839509	85.11912541	431951
5	-1.018589327	0.36548087	120.3055431	416237
6	-1.024301295	0.362039931	117.400173	403946
7	-1.268517165	0.46158093	147.2360086	390869
8	-0.926111907	0.429499491	134.9514917	379253
9	-0.345378227	0.074466527	23.08613627	369215
10	-0.562055342	0.197698727	60.40675345	358643
11	-0.88808892	0.33603274	101.2165324	348529
12	-0.325775445	0.143899462	42.82453389	340225
13	-0.42164673	0.204894026	60.28030618	332500
14	-0.424273829	0.15590397	45.2967324	324279
15	-0.54853456	0.239534194	68.66418824	315664
16	-0.301133365	0.075040559	21.24141239	307804
17	-0.266106236	0.063168846	17.66783311	300511
18	-0.478598559	0.142411706	39.30217942	292578
19	-0.342197989	0.176669872	48.21389913	286101
20	-0.347000772	0.161187188	43.48176743	279546

}{\neg}

\pgfplotstableread{
x  mean
1	-0.544113253
2	-1.239954782
3	0.0500611
4	-0.100234099
5	0.034166265
6	-0.243276167
7	-0.627194918
8	0.412973237
9	0.378859107
10	0.297335136
11	-0.220747464
12	0.264094586
13	-0.028124956
14	0.158265505
15	-0.188096931
16	0.252114967
17	0.321254954
18	-0.05172815
19	-0.058954578
20	-0.085661715

}{\mid}

\begin{tikzpicture}

\pgfplotsset{
every axis legend/.append style={
at={(0.5,-0.650)},
anchor=north
},
}
 
 \begin{groupplot}[
    group style={%
        group size=1 by 2,%
        x descriptions at=edge bottom,%
        vertical sep=0pt,%
    },
    clip=true,
    clip mode=individual,
    width=1.75in,
    xmin=0,
    xmax=20,
    legend columns=3,
    legend style={draw=none}
    %ymode=log,
    ]
    \nextgroupplot[
        title = Team Size 3-5,
        xticklabels={,,}, 
        height=1.5in, 
        ymax=2.2,
        ymin=-4,
        y label style={at={(axis description cs:0.08,.5)}},
        yticklabels={,,},
        ]
    \addplot [stack plots=y, fill=none, draw=none, forget plot]   table [x=x, y expr=(\thisrow{mean} + \thisrow{ci})]   {\pos} \closedcycle;
    \addplot [stack plots=y, fill=gray!70, opacity=0.6, draw opacity=1, thin, smooth, area legend]   table [x=x, y expr=(\thisrow{mean} - \thisrow{ci}) - (\thisrow{mean} + \thisrow{ci}) ]   {\pos} \closedcycle;
    \addplot [stack plots=y, stack dir=minus, forget plot, draw=none] table [x=x, y expr = (\thisrow{mean} - \thisrow{ci})] {\pos};

    \addplot [stack plots=y, fill=none, draw=none, forget plot]   table [x=x, y expr=(\thisrow{mean} + \thisrow{ci})]   {\neg} \closedcycle;
    \addplot [stack plots=y, fill=gray!70, opacity=0.6, draw opacity=1, thin, smooth, area legend]   table [x=x,y expr=(\thisrow{mean} - \thisrow{ci}) - (\thisrow{mean} + \thisrow{ci}) ]   {\neg} \closedcycle;
    \addplot [stack plots=y, stack dir=minus, forget plot, draw=none] table [x=x, y expr = (\thisrow{mean} - \thisrow{ci})] {\neg};
    
    \addplot [stack plots=false, green, thick, smooth]  table [x=x, y=mean]
    {\pos};
    \addplot [stack plots=false, red, thick, smooth]  table [x=x, y=mean]   
    {\neg};
    
    \draw[ultra thin] (axis cs:\pgfkeysvalueof{/pgfplots/xmin},0) -- (axis cs:\pgfkeysvalueof{/pgfplots/xmax},0);
    
    \addplot [stack plots=false, black, thick, smooth]  table [x=x, y=mean]   
    {\mid};
    \nextgroupplot[
        ybar, 
        bar width=2pt, 
        ymode=log,
        height=0.95in, 
        ymax=1000000,
        ymin=100000,
        yticklabels={,,},
        xticklabels={,0,5,10,15},
    ]
    \addplot []  table [x=x, y=len]   {\pos};
    %\node[anchor=north east] at (rel axis cs:-0.05,-0.05) {\textbf{(c)}};
        
    \end{groupplot}
\end{tikzpicture}
    \end{minipage}
    \begin{minipage}{0.15\textwidth}
        \pgfplotstableread{
x  mean  ci std len
1	1.050850567	0.282333165	57.54468068	159584
2	0.750941434	0.098050972	19.7043593	155140
3	0.649013132	0.204355948	40.45898851	150577
4	0.553873642	0.0950642	18.60978549	147215
5	0.636505815	0.124343764	24.04350965	143632
6	0.64472753	0.16657971	31.89329309	140818
7	0.583553948	0.186334703	35.36437309	138372
8	0.730924484	0.304116402	57.20694271	135932
9	0.976807074	0.299351178	55.79806588	133469
10	0.99260832	1.360276141	251.8535629	131688
11	0.820900282	0.990993443	182.3625256	130087
12	0.799996505	0.377335269	68.86906641	127967
13	0.302849204	0.094573285	17.16167896	126499
14	0.561322115	0.192976047	34.78718491	124835
15	0.650074703	0.272893515	48.85595475	123127
16	0.331428621	0.073441384	13.05466031	121382
17	1.102919976	0.585756189	103.1548223	119138
18	0.42503473	0.239233942	41.82960838	117443
19	0.448454824	0.207425178	35.96648037	115499
20	0.494984841	0.233972036	40.32965296	114137

}{\pos}
%TW adjusted to remove outlier in #2
\pgfplotstableread{
x  mean  ci std len
%0	0	0	0	0.0	0	0	0	-0.0	0.0	0.0
1	-2.241859104	0.456181256	92.97811233	159584
2	-1.61134390	    0.699801245	1171.150839	155140 
3	-1.216608564	0.364864707	72.23698227	150577
4	-0.838829039	0.19524087	38.22038917	147215
5	-0.926458896	0.236827624	45.79375028	143632
6	-0.589897163	0.125978828	24.11986233	140818
7	-0.650568356	0.359070333	68.14778484	138372
8	-0.74024272	0.590623431	111.1014092	135932
9	-0.363682035	0.188532949	35.14191581	133469
10	-0.349304955	0.123288772	22.8267743	131688
11	-0.545289333	0.245796252	45.23140452	130087
12	-0.532926179	0.339213063	61.91122022	127967
13	-0.549146811	1.00027142	181.5135957	126499
14	-0.398141595	0.152384289	27.46983651	124835
15	-0.431373577	0.442970462	79.30472375	123127
16	-1.165552672	1.482325016	263.492441	121382
17	-0.611259644	0.248352408	43.73619784	119138
18	-0.403864468	0.44033666	76.99204319	117443
19	-0.488043127	0.378021395	65.54700458	115499
20	-0.511196736	0.501939266	86.51904188	114137

}{\neg}
%TW adjusted to remove outlier earlier in #2 below
\pgfplotstableread{
x  mean
1	-1.191008536
2	-0.8604 
3	-0.567595432
4	-0.284955397
5	-0.289953081
6	0.054830367
7	-0.067014408
8	-0.009318235
9	0.613125038
10	0.643303365
11	0.275610949
12	0.267070326
13	-0.246297607
14	0.16318052
15	0.218701126
16	-0.834124051
17	0.491660332
18	0.021170262
19	-0.039588303
20	-0.016211895

}{\mid}

\begin{tikzpicture}

\pgfplotsset{
every axis legend/.append style={
at={(0.5,-0.650)},
anchor=north
},
}
 
 \begin{groupplot}[
    group style={%
        group size=1 by 2,%
        x descriptions at=edge bottom,%
        vertical sep=0pt,%
    },
    clip=true,
    clip mode=individual,
    width=1.75in,
    xmin=0,
    xmax=20,
    legend columns=3,
    legend style={draw=none}
    %ymode=log,
    ]
    \nextgroupplot[
        title = Team Size 6-9,
        xticklabels={,,}, 
        height=1.5in, 
        ymax=2.2,
        ymin=-4,
        y label style={at={(axis description cs:0.08,.5)}},
        yticklabels={,,},
        ]
    \addplot [stack plots=y, fill=none, draw=none, forget plot]   table [x=x, y expr=(\thisrow{mean} + \thisrow{ci})]   {\pos} \closedcycle;
    \addplot [stack plots=y, fill=gray!70, opacity=0.6, draw opacity=1, thin, smooth, area legend]   table [x=x, y expr=(\thisrow{mean} - \thisrow{ci}) - (\thisrow{mean} + \thisrow{ci}) ]   {\pos} \closedcycle;
    \addplot [stack plots=y, stack dir=minus, forget plot, draw=none] table [x=x, y expr = (\thisrow{mean} - \thisrow{ci})] {\pos};

    \addplot [stack plots=y, fill=none, draw=none, forget plot]   table [x=x, y expr=(\thisrow{mean} + \thisrow{ci})]   {\neg} \closedcycle;
    \addplot [stack plots=y, fill=gray!70, opacity=0.6, draw opacity=1, thin, smooth, area legend]   table [x=x,y expr=(\thisrow{mean} - \thisrow{ci}) - (\thisrow{mean} + \thisrow{ci}) ]   {\neg} \closedcycle;
    \addplot [stack plots=y, stack dir=minus, forget plot, draw=none] table [x=x, y expr = (\thisrow{mean} - \thisrow{ci})] {\neg};
    
    \addplot [stack plots=false, green, thick, smooth]  table [x=x, y=mean]
    {\pos};
    \addplot [stack plots=false, red, thick, smooth]  table [x=x, y=mean]   
    {\neg};
    
    \draw[ultra thin] (axis cs:\pgfkeysvalueof{/pgfplots/xmin},0) -- (axis cs:\pgfkeysvalueof{/pgfplots/xmax},0);
    
    \addplot [stack plots=false, black, thick, smooth]  table [x=x, y=mean]   
    {\mid};\legend{}
    
    \nextgroupplot[
        ybar, 
        bar width=2pt, 
        ymode=log,
        height=0.95in, 
        ymax=1000000,
        ymin=100000,
        yticklabels={,,},
        xticklabels={,0,5,10,15},
    ]
    \addplot []  table [x=x, y=len]   {\pos};
    %\node[anchor=north east] at (rel axis cs:-0.05,-0.05) {\textbf{(c)}};
    \end{groupplot}
\end{tikzpicture}
    \end{minipage}
    \begin{minipage}{0.15\textwidth}
        \pgfplotstableread{
x  mean  ci std len
1	1.637586069	0.351649057	111.6758934	387435
2	0.65854964	0.163973665	51.04189465	372224
3	0.341643807	0.029071153	8.93317118	362732
4	0.373245521	0.047925554	14.5680412	354950
5	0.473596635	0.05171692	15.57389312	348360
6	0.327974506	0.0617564	18.44141882	342550
7	0.426191882	0.21921352	65.01175089	337869
8	0.542986011	0.659301851	194.1221087	333028
9	0.292023865	0.082856886	24.22818173	328461
10	0.22722777	0.042008573	12.2200786	325066
11	0.258817848	0.064472764	18.64374935	321228
12	0.236628142	0.038664269	11.11580227	317513
13	0.195349703	0.028153691	8.053347104	314327
14	0.160452114	0.0272595	7.754491512	310864
15	0.210005149	0.043776243	12.39117273	307785
16	0.227892472	0.081097213	22.82593061	304330
17	0.298182178	0.037582866	10.51430997	300664
18	0.331939447	0.040715706	11.27226628	294441
19	0.228346273	0.141564887	38.95480948	290878
20	0.258880391	0.065428701	17.90339807	287630

}{\pos}

\pgfplotstableread{
x  mean  ci std len
%0	0	0	0	0.0	0	0	0	-0.0	0.0	0.0
1	-1.292118407	0.260338105	82.67757248	387435
2	-0.888578759	0.164608251	51.23942915	372224
3	-0.393580752	0.052277554	16.06418334	362732
4	-0.505307436	0.169555814	51.5402719	354950
5	-0.455338416	0.069717101	20.99441887	348360
6	-0.518664021	0.156759659	46.8108653	342550
7	-0.23761865	0.080257004	23.8016723	337869
8	-0.403392744	0.654759317	192.7846238	333028
9	-0.128770403	0.016772387	4.904413624	328461
10	-0.130079444	0.020998326	6.108305346	325066
11	-0.242047573	0.086758298	25.08811268	321228
12	-0.165817729	0.03820459	10.98364665	317513
13	-0.113360077	0.017988988	5.145739632	314327
14	-0.201265116	0.084573872	24.05867191	310864
15	-0.086854897	0.013227282	3.744074978	307785
16	-0.308135256	0.122407248	34.45321068	304330
17	-0.209026277	0.033888981	9.480896203	300664
18	-0.294902302	0.038758017	10.73027412	294441
19	-0.119439847	0.044500969	12.2454572	290878
20	-0.220033861	0.074300914	20.33112113	287630

}{\neg}

\pgfplotstableread{
x  mean
1	0.345467662
2	-0.230029119
3	-0.051936945
4	-0.132061915
5	0.018258219
6	-0.190689516
7	0.188573232
8	0.139593266
9	0.163253462
10	0.097148326
11	0.016770275
12	0.070810413
13	0.081989627
14	-0.040813002
15	0.123150252
16	-0.080242784
17	0.089155901
18	0.037037145
19	0.108906427
20	0.038846531

}{\mid}

\begin{tikzpicture}

\pgfplotsset{
every axis legend/.append style={
at={(0.5,-0.650)},
anchor=north
},
}
 
 \begin{groupplot}[
    group style={%
        group size=1 by 2,%
        x descriptions at=edge bottom,%
        vertical sep=0pt,%
    },
    clip=true,
    clip mode=individual,
    width=1.75in,
    xmin=0,
    xmax=20,
    legend columns=3,
    legend style={draw=none}
    %ymode=log,
    ]
    \nextgroupplot[
        title = Team Size $\ge$10,
        xticklabels={,,}, 
        height=1.5in, 
        ymax=2.2,
        ymin=-4,
        y label style={at={(axis description cs:0.08,.5)}},
        yticklabels={,,},
        ]
    \addplot [stack plots=y, fill=none, draw=none, forget plot]   table [x=x, y expr=(\thisrow{mean} + \thisrow{ci})]   {\pos} \closedcycle;
    \addplot [stack plots=y, fill=gray!70, opacity=0.6, draw opacity=1, thin, smooth, area legend]   table [x=x, y expr=(\thisrow{mean} - \thisrow{ci}) - (\thisrow{mean} + \thisrow{ci}) ]   {\pos} \closedcycle;
    \addplot [stack plots=y, stack dir=minus, forget plot, draw=none] table [x=x, y expr = (\thisrow{mean} - \thisrow{ci})] {\pos};

    \addplot [stack plots=y, fill=none, draw=none, forget plot]   table [x=x, y expr=(\thisrow{mean} + \thisrow{ci})]   {\neg} \closedcycle;
    \addplot [stack plots=y, fill=gray!70, opacity=0.6, draw opacity=1, thin, smooth, area legend]   table [x=x,y expr=(\thisrow{mean} - \thisrow{ci}) - (\thisrow{mean} + \thisrow{ci}) ]   {\neg} \closedcycle;
    \addplot [stack plots=y, stack dir=minus, forget plot, draw=none] table [x=x, y expr = (\thisrow{mean} - \thisrow{ci})] {\neg};
    
    \addplot [stack plots=false, green, thick, smooth]  table [x=x, y=mean]
    {\pos};
    \addplot [stack plots=false, red, thick, smooth]  table [x=x, y=mean]   
    {\neg};
    
    \draw[ultra thin] (axis cs:\pgfkeysvalueof{/pgfplots/xmin},0) -- (axis cs:\pgfkeysvalueof{/pgfplots/xmax},0);
    
    \addplot [stack plots=false, black, thick, smooth]  table [x=x, y=mean]   
    {\mid};\legend{}
    
    \nextgroupplot[
        ybar, 
        bar width=2pt, 
        ymode=log,
        height=0.95in, 
        ymax=1000000,
        ymin=100000,
        yticklabels={,,},
        xticklabels={,0,5,10,15,20},
    ]
    \addplot []  table [x=x, y=len]   {\pos};
    %\node[anchor=north east] at (rel axis cs:-0.05,-0.05) {\textbf{(c)}};
    \end{groupplot}
\end{tikzpicture}
    \end{minipage}

\begin{tikzpicture}
    \begin{customlegend}[legend columns=-1,
      legend style={
        draw=none,
        column sep=1ex,
      },
    legend entries={Additions, Deletions, Net Use}]
    \addlegendimage{green, mark=*}
    \addlegendimage{red, mark=*}
    \addlegendimage{black, mark=*}
    \end{customlegend}
\end{tikzpicture}

    \caption{Library additions, deletions and net-usage (in LOC) after the adoption event. }
    \label{fig:afteradoptionteamsize}
\end{figure*}

\begin{table}[]
    \centering
    \caption{Statistics surrounding newly adopted library $\ell$}
    \begin{tabular}{c|c}
         Avg LOC that reference $\ell$ & 31.34 \\
         Median LOC that reference $\ell$ & 4 \\ \hline
         Avg insert LOC after 1st adoption to ref $\ell$ & 2.09 \\
         Avg deleted LOC after 1st adoption to ref $\ell$ & 1.62 \\
    \end{tabular}
    \label{tab:adoption_stats}
\end{table}

Table \ref{tab:adoption_stats} shows there is a wide gap between the average LOC and median lines of code that represent a library $\ell$, indicating skew caused by large commits. This matches with our analysis which showed earlier than many of the statistics surrounding commits follow a power law distribution. Additionally, average LOC drops quickly after the first commit. Fig.~\ref{fig:afteradoptionteamsize} shows the average direct use of an adopted library in lines added and deleted (in green and red respectively) as well as the net change, \ie, insertions minus deletions (in black). 

After the initial commit, we find that most of the following commits have only a small positive net productivity. We also find that the volume of activity of lines of code referencing $\ell$ in Fig.~\ref{fig:adopt_per_commit} tends towards zero rather quickly after the adoption. This indicates that, on average, the activity surrounding the adoption of a library is brief and oftentimes contradicted quickly. Recall from Fig.~\ref{fig:adopt_dist} that most repositories only adopt a few libraries, with more than half adopting 10 or fewer libraries. Therefore, we can safely deduce that in most repositories, when adoptions occur, they occur early in a repository's history.

\paragraph{Stack Overflow}

The popularity of software-centric question-and-answer Web sites has spread rapidly. Stack Overflow was created in 2008~\cite{anderson2012discovering}, which is the same year that GitHub was launched~\cite{yan2017empirical}. Because these resources have grown in popularity together, we expect that they have influenced each other in some way. Much of the previous work in this area has focused on understanding user behaviors across Stack Overflow and GitHub~\cite{xiong2017mining,silvestri2015linking}, modelling user interest~\cite{lee2017GitHub}, or mining for code that users have imported from Stack Overflow posts into GitHub projects~\cite{gharehyazie2017some}. In other cases, researchers aim to leverage posts and their answers in order to build tools that aid software development~\cite{ponzanelli2014mining,reiss2009semantics}. Further research needs to be cone to understand data flows from Stack Overflow to GitHub, and vice-versa.

\begin{figure}[t]
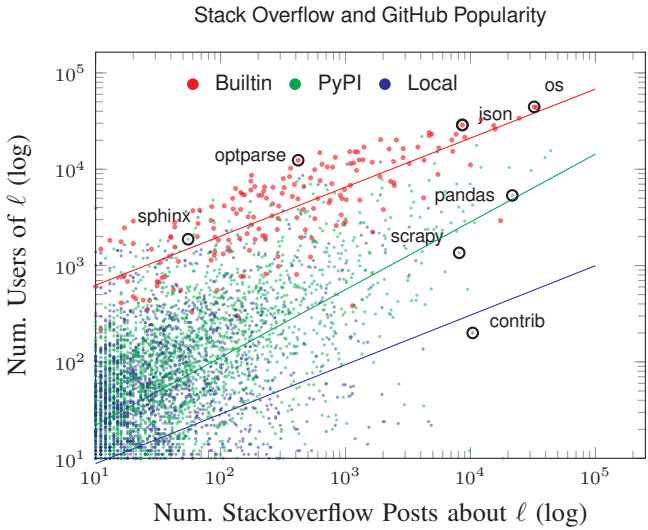

    \centering
        \include{./figs/stackoverflow_users}
    \caption{Number of users of $\ell$ in GitHub dataset as a function of the number of Stack Overflow posts about $\ell$, showing that each library type has a statistically significant positive correlation.}
    \label{fig:StackOverflow_users}
\end{figure}
\nop{
\begin{figure}[t]
    \centering
        \pgfplotstableread{
x	one	hundred	thousand	tenthousand

0 1.0 1.0 1.0 1.0
1 1.1 1.1810730253353205 1.16 1.2222222222222223
2 1.2613636363636362 1.25 1.16 1.34375
3 1.2 1.2666666666666666 1.4705882352941178 1.5
4 1.3333333333333333 1.4962686567164178 1.6 1.6478537360890302
5 1.434782608695652 1.5555555555555556 1.75 1.75
6 1.4 1.6221774193548386 1.8333333333333333 2.0
7 1.4848484848484849 1.7951388888888888 2.0 2.0
8 1.6862745098039216 2.0 2.0 2.0714285714285716
9 1.7945478723404256 2.0 2.1099656357388317 2.2
10 1.725 2.0 2.2981481481481483 2.4
11 1.8888888888888888 2.1 2.4142156862745097 2.5
12 1.75 2.0 2.5454545454545454 2.5454545454545454
13 1.7471264367816093 2.375 2.619047619047619 2.727272727272727
14 1.9130434782608696 2.4 2.6576576576576576 2.923076923076923
15 2.3229813664596275 2.2962962962962963 2.9166666666666665 2.888888888888889
16 2.4096153846153845 2.8 2.8 3.0
17 2.1176470588235294 2.7216386554621845 2.8095238095238093 3.0
18 2.0 2.3125 2.8117647058823527 3.111111111111111
19 2.8 2.8320802005012533 3.1225 3.2264957264957266
20 2.5 2.655952380952381 3.111111111111111 3.3333333333333335
21 2.1145940390544706 3.0115384615384615 3.32 3.6666666666666665
22 2.2564102564102564 3.2666666666666666 3.327777777777778 3.5883116883116886
23 2.6195054945054945 3.0 3.453463203463204 3.6
24 2.5774886877828056 2.54296875 3.3301282051282053 3.6704545454545454
25 2.6904761904761907 3.641025641025641 3.3333333333333335 3.7247596153846154
26 2.75 3.3095238095238093 3.5 3.8578595317725752
27 2.1816418875242403 2.5 3.3333333333333335 3.8944444444444444
28 3.772222222222222 3.466666666666667 3.5 4.11437908496732
29 3.0 3.2 3.8157894736842106 4.0
30 2.3076923076923075 4.182539682539683 4.267857142857142 4.318181818181818
31 3.0625 4.587606837606838 3.8958333333333335 4.291666666666667
32 3.388888888888889 3.357142857142857 4.0 4.5
33 3.3375 3.5892857142857144 4.0 4.5
34 4.375 4.5 4.115079365079366 4.294117647058823
35 3.5625 4.270833333333333 4.086904761904762 4.333333333333333
36 3.201530612244898 3.5789473684210527 4.333333333333333 4.4
37 2.532488114104596 4.8 4.481481481481482 4.25
38 3.3333333333333335 4.923611111111111 4.25 4.4
39 3.4615384615384617 3.6666666666666665 4.4 4.67816091954023
40 3.4705882352941178 3.466666666666667 4.192028985507246 4.666666666666667
41 3.7080745341614905 3.6 4.166666666666667 4.666666666666667
42 2.5 4.060855263157895 4.552727272727273 5.0
43 4.0 4.4 4.571428571428571 5.25
44 4.0 4.107142857142857 4.733333333333333 5.111111111111111
45 2.7717948717948717 3.5 4.980769230769231 5.0
46 3.472463768115942 4.8273809523809526 4.895833333333333 5.277777777777778
47 3.891304347826087 5.455882352941177 4.809090909090909 5.58578431372549
48 3.6666666666666665 3.9444444444444446 4.881944444444445 5.236111111111111
49 3.3333333333333335 3.7142857142857144 4.8076923076923075 5.25
50 3.81055900621118 5.4451706608569355 4.75 5.6201923076923075

}{\table}

\begin{tikzpicture}

\pgfplotsset{
every axis legend/.append style={
at={(0.5, -0.30)}, 
anchor=north
},
}

\begin{axis}[
    legend style={draw=none}, 
    title = Numpy Adoption Rate by Number of Posts on Stack Overflow, 
    width=3.50in,
    height=2.250in, 
    xlabel = {Numpy Adoption Rate}, 
    ylabel = {median \% change (LOC$_\ell$)}, 
    legend columns=4
    ]
    \addplot [blue,  thick]  table [x=x,  y=one]   {\table};
    \addplot [blue!50!green,  thick]  table [x=x,  y=hundred]   {\table};
    \addplot [green,  thick]  table [x=x,  y=thousand]   {\table};
    \addplot [orange,  thick]  table [x=x,  y=tenthousand]   {\table};
    
    \legend{0-100, 101-1000, 1001-10000, 10000+}
    \end{axis}
\end{tikzpicture}
    \caption{Here we see how the \texttt{numpy} library adoption has become faster as the number of posts that appear on Stack Overflow have increased.}
    \label{fig:numpy_Stack Overflow}
\end{figure}

\begin{figure}[t]
    \centering
        \include{./figs/numpy_overtime}
    \caption{Here we see how the \texttt{numpy} library adoption has become faster as the number of posts that appear on Stack Overflow have increased.}
    \label{fig:numpy_Stack Overflow}
\end{figure}}

\begin{figure*}[t]
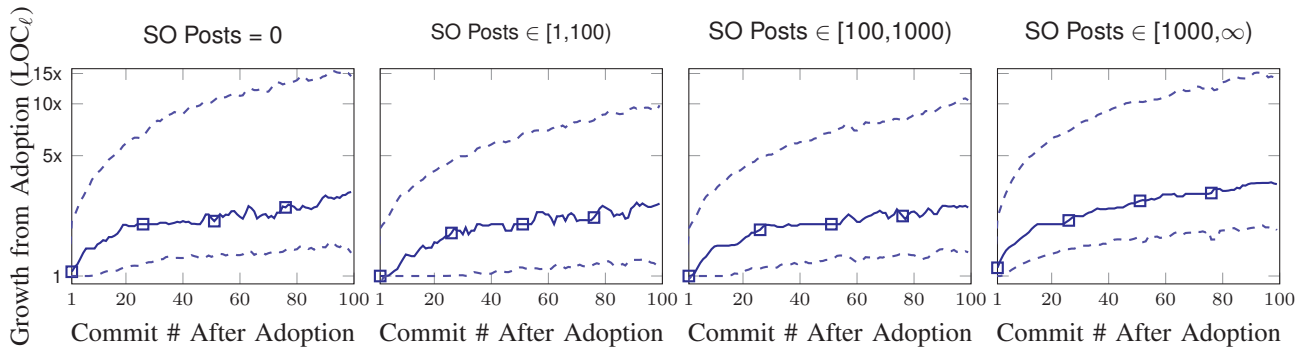

    \centering
        \include{./figs/adoptions_by_stackoverflow}
    \caption{Growth of library usage after adoption grouped by Stack Overflow usage. Q1, Median, and Q3 growth in lines of code referencing $\ell$ are represented from bottom to top respectively.}
    \label{fig:StackOverflow}
\end{figure*}

We plot the number of users of $\ell$ by the mean average number of Stack Overflow posts (across all adoption times) in Fig.~\ref{fig:StackOverflow_users} that existed when $\ell$ was referenced. This illustration also groups libraries that are (1) included in PyPi, the default library repository used by the pip installer, (2) part of Python's standard suite of libraries, \eg, os, json, time, and (3) all other libraries. We observe a strong positive correlation between the number of library users and the number of Stack Overflow mentions for standard libraries ($R^2$=0.625, $p<$0.001) and PyPi libraries ($R^2$=0.410, $p<$0.001). There is a small positive correlation between usage of unknown libraries and Stack Overflow posts ($R^2$=0.08, $p<$0.001), most likely due to individuals naming libraries like words and phrases that also happen to appear on Stack Overflow, or perhaps users sharing GitHub repositories that have not made it to PyPi.

Exemplar libraries are called out in Fig.~\ref{fig:StackOverflow_users}. For example, the standard libraries \texttt{optparse}, \texttt{json}, and \texttt{os} indicate some of the most widely used libraries on GitHub. The \texttt{sphinx}, \texttt{scrapy}, and \texttt{pandas} libraries represent three libraries found on PyPi. The \texttt{sphinx} library is used to create source code documentation files; it is relatively popular on GitHub but has only a few dozen posts on Stack Overflow. This seems to indicate that users have few questions about this library relative to its use (perhaps the library that produces source code documentation is well-documented!). Conversely, the \texttt{scrapy} library, which is used to crawl the Web, and the \texttt{pandas} library, which is used for data analysis, have many questions, potentially indicating that the library is complicated to use. Despite being rather popular on Stack Overflow and GitHub, the \texttt{contrib} ``library'' is not found in the standard Python libraries nor PyPi. This is a bit of a misnomer because the use of a ``contrib'' folder/module is a standard way to encourage open source developers to \emph{contrib}ute to various projects. As a result, the contrib module is indicated as a common library simply because of its use as a naming convention in many distinct projects, so we see that it is mentioned quite frequently on Stack Overflow as a ``local'' library, but the functions defined in local ``contrib'' libraries across Github would yield very different functions since each user is writing their own ``contrib'' library for different purposes.

Our next task is to understand what differences, if any, exist in the net productivity of these various libraries. To help understand the dynamics of library adoption, we calculate the median percentage growth (in LOC) of an adopted library, \ie, the change in the number of added lines containing $\ell$ in a commit minus the number of deleted lines containing $\ell$ in a commit for the first 100 commits after the adoption. This provides us with a simple way to compare growth across teams which are different sizes.

Formally, we compute the growth of a library $\ell$ within a project as follows. If $x=0$, then let $y_{x} = 1$; otherwise
$$y_{x} = y_{x-1} \left(\frac{\sum\limits_{i=0}^{x-1}{(n_i)} + n_x}{\sum\limits_{i=0}^{x-1}{(n_i)}}\right),$$
where $n_i$ is the number of changed lines of code (net additions minus deletions) in commit $i$ that contain $\ell$. From this equation $y_x$ contains the percentage change at commit $x$ relative to the adoption event ($x=0$).

Consider as an example the following series of commits $n$ = [+2, +1, +4, -1]. Here, the adoption commit ($x=0$) introduces two lines of code that reference $\ell$. We set $y_{0} = 1$. The next commit contains a net change of +1, rendering $y_1 = 1((2 + 1)/2) = 1.5$. In other words, in the second commit the number of lines of code referencing $\ell$ within this example project grew such that it is now 150\% of its original size. The next commit contains a net change of +3, rendering $y_2 = 1.5((3 + 4)/3) = 3.5$ indicating that after the third commit the use of $\ell$ within this example project grew to be 350\% of its size over the initial commit, \ie, from 2 lines to 7. The final commit contains a net change of -1, rendering $y_3 = 3.5((7 - 1 )/7) = 3$. This normalization of median growth rate helps us compare larger teams to smaller ones.

We plot the median growth (in LOC referencing $\ell$) as a function of the number of commits after the adoption in Fig.~\ref{fig:StackOverflow}. Note that the adoption commit is not shown; instead, each commit either net-adds or net-subtracts from the initial adoption. Columns represent four groups of Stack Overflow mention sizes: no mention, between 1 and 100, 100 and 1000, and greater than 1000 from left to right respectively. Within each plot, solid lines represent the median, and the dashed lines on top and bottom represent the 3rd and 1st quartiles respectively. 

%The three rows in Fig.~\ref{fig:stackoverflow} represents Builtin, PyPi, or Private libraries from top to bottom respectively. 

%The third quartile line is clipped in the figure in order to clearly illustrate the median. There were no Builtin libraries that did not appear on Stack Overflow, so the respective plot is missing. Likewise, there are very few libraries with 1000 or more mentions on Stack Overflow, so the respective plot is not shown.

We observe that the use of an adopted library has a complicated relationship with its popularity on Stack Overflow. The primary distinction is in the growth rates for libraries with more than 1000 Stack Overflow posts. 100 commits after the adoption, the adoption of a highly mentioned library on Stack Overflow will have approximately 350\% growth (on average) compared to only 250\% growth for less mentioned libraries. We can assume that having over 1000 Stack Overflow posts means that the library is highly successful, and highly popular. Over time, it might be easier for users to find new resources about the library online - and add more functionality as the library becomes fully integrated into the project.

%Across the different library types, \ie, Builtin, PyPi, and local, show similar behavior. The primary distinction is in the growth rates for libraries with more than 1000 Stack Overflow posts. 100 commits after the adoption, the adoption of a highly mentioned library on Stack Overflow will have approximately 350\% growth (on average) compared to only 250\% growth for less mentioned libraries. 

When a library does not appear on Stack Overflow, the growth rate is similar to libraries that have over 1000 posts. Libraries that do not appear at all in Stack Overflow mostly consist of libraries that were written by developers who are also the authors committing the library to the repository. This may explain why growth is large in unknown libraries -- the adopters know how to use the library because they wrote it. We could also propose that programmers who are using libraries for which there are no online resources available might be more experienced than those that are using more popular libraries, so their growth rate is faster.

%Stack Overflow has emerged as a helpful tool for programmers to share their knowledge to others.  We expect that popular Stack Overflow posts will continue to help new users understand how to use new libraries more fully, and that new libraries will need to use Stack Overflow to grow in popularity.

\paragraph{Project Team Size}

Next, we investigate differences in library adoptions as a function of team size. Because library adoptions occur from the perspective of a project, studying how various team sizes adopt libraries is important as we attempt to understand how teams form and work together. Researchers have studied GitHub previously for its team formation attributes. Git and GitHub directly store how team members collaborate and the types of activities that they perform~\cite{middleton2018contributions}. For example, researchers have found that diversity and team makeup have a significant impact on the productivity of GitHub teams~\cite{vasilescu2015data}, and larger teams tend to process more pull requests on average~\cite{vasilescu2015quality}.

\begin{figure}[t]
\begin{minipage}[t]{0.46\linewidth}
\centering
    \pgfplotstableread{
Size    Number
1	0.590837131
2	0.241474811
3	0.072450147
4	0.029347529
5	0.01516488
6	0.009294604
7	0.005989685
8	0.004329521
9	0.003004472
10	0.002438245
11	0.001952907
12	0.001629348
13	0.001413643
14	0.001194085
15	0.001009194
16	0.000855119
17	0.000697192
18	0.000620154
19	0.000624006
20	0.000516153
21	0.000543116
22	0.000473782
23	0.000516153
24	0.000377485
25	0.000323559
26	0.000365929
27	0.000342818
28	0.00034667
29	0.000238817
30	0.000246521
31	0.000342818
32	0.000215706
33	0.000223409
34	0.000227261
35	0.000227261
36	0.000215706
37	0.000234965
38	0.000154075
39	0.000157927
40	0.000165631
41	0.000157927
42	0.000146372
43	0.00014252
44	9.24453E-05
45	0.000130964
46	0.00014252
47	0.000119409
48	0.000154075
49	0.000119409
50	0.000130964
51	0.000146372
52	0.000169483
53	0.000100149
54	0.000157927
55	0.000177187
56	0.000130964
57	0.00014252
58	0.000104001
59	7.70377E-05
60	0.000154075
61	0.00012326
62	0.000130964
63	0.000146372
64	0.00012326
65	0.000154075
66	8.08896E-05
67	0.000119409
68	8.47415E-05
69	0.00014252
70	0.000107853
71	4.23708E-05
72	7.31859E-05
73	7.31859E-05
74	0.000115557
75	0.000104001
76	0.000104001
77	7.31859E-05
78	7.70377E-05
79	7.70377E-05
80	6.54821E-05
81	4.23708E-05
82	4.23708E-05
83	6.54821E-05
84	0.000173335
85	9.62972E-05
86	6.9334E-05
87	5.00745E-05
88	8.08896E-05
89	5.39264E-05
90	4.23708E-05
91	7.31859E-05
92	2.31113E-05
93	2.31113E-05
94	4.62226E-05
95	3.08151E-05
96	4.62226E-05
97	5.39264E-05
98	7.70377E-05
99	6.54821E-05
100	8.08896E-05
101	3.08151E-05
102	3.08151E-05
103	5.39264E-05
104	2.69632E-05
105	2.69632E-05
106	4.23708E-05
107	7.70377E-06
108	1.92594E-05
109	3.85189E-05
110	2.31113E-05
111	3.4667E-05
112	1.92594E-05
113	3.4667E-05
114	2.31113E-05
115	1.92594E-05
116	1.54075E-05
117	2.31113E-05
118	3.08151E-05
119	1.54075E-05
120	7.70377E-06
121	1.92594E-05
122	1.92594E-05
123	3.08151E-05
124	7.70377E-06
125	7.70377E-06
126	3.08151E-05
127	2.31113E-05
128	7.70377E-06
129	3.4667E-05
130	2.69632E-05
131	1.54075E-05
132	2.69632E-05
133	3.85189E-05
134	1.54075E-05
135	1.54075E-05
136	7.70377E-06
137	1.15557E-05
138	1.54075E-05
139	2.31113E-05
140	1.54075E-05
141	3.4667E-05
142	7.70377E-06
143	1.15557E-05
144	7.70377E-06
145	2.31113E-05
146	1.54075E-05
147	1.15557E-05
148	2.69632E-05
149	1.15557E-05
150	1.92594E-05
151	1.54075E-05
152	1.92594E-05
153	1.54075E-05
154	1.15557E-05
155	7.70377E-06
156	1.54075E-05
157	1.92594E-05
158	1.15557E-05
159	3.08151E-05
160	1.54075E-05
161	3.85189E-06
162	1.54075E-05
163	1.15557E-05
164	7.70377E-06
165	1.92594E-05
166	1.15557E-05
167	2.31113E-05
169	1.54075E-05
171	2.69632E-05
172	7.70377E-06
173	7.70377E-06
174	1.15557E-05
175	1.15557E-05
176	1.54075E-05
177	1.92594E-05
179	1.15557E-05
180	1.92594E-05
181	7.70377E-06
182	1.15557E-05
183	3.85189E-06
184	1.15557E-05
186	7.70377E-06
187	7.70377E-06
188	1.15557E-05
189	2.31113E-05
190	1.54075E-05
191	2.31113E-05
192	1.15557E-05
193	1.15557E-05
194	3.85189E-06
195	1.92594E-05
196	3.85189E-06
198	1.15557E-05
199	7.70377E-06
200	7.70377E-06
201	1.54075E-05
202	7.70377E-06
203	3.85189E-06
204	1.15557E-05
205	3.85189E-06
206	3.85189E-06
209	3.85189E-06
210	1.54075E-05
211	1.15557E-05
212	7.70377E-06
213	3.85189E-06
214	7.70377E-06
215	1.54075E-05
216	1.15557E-05
220	7.70377E-06
221	7.70377E-06
222	3.85189E-06
223	3.85189E-06
224	7.70377E-06
225	1.15557E-05
226	7.70377E-06
227	3.85189E-06
228	1.15557E-05
229	1.54075E-05
230	1.15557E-05
231	1.54075E-05
232	1.54075E-05
233	1.15557E-05
234	1.15557E-05
235	3.85189E-06
236	7.70377E-06
237	7.70377E-06
238	3.85189E-06
240	1.15557E-05
242	3.85189E-06
243	1.15557E-05
244	7.70377E-06
245	3.85189E-06
246	3.85189E-06
247	3.85189E-06
248	3.85189E-06
249	3.85189E-06
250	1.15557E-05
251	7.70377E-06
252	3.85189E-06
253	1.15557E-05
254	3.85189E-06
255	3.85189E-06
256	3.85189E-06
258	3.85189E-06
259	7.70377E-06
260	3.85189E-06
261	7.70377E-06
262	7.70377E-06
263	3.85189E-06
265	3.85189E-06
266	3.85189E-06
267	3.85189E-06
268	3.85189E-06
269	1.15557E-05
270	7.70377E-06
271	7.70377E-06
272	3.85189E-06
273	1.15557E-05
276	7.70377E-06
277	1.54075E-05
283	3.85189E-06
285	1.15557E-05
286	3.85189E-06
287	1.15557E-05
288	3.85189E-06
289	3.85189E-06
290	3.85189E-06
291	3.85189E-06
292	3.85189E-06
294	3.85189E-06
295	3.85189E-06
297	3.85189E-06
298	3.85189E-06
300	3.85189E-06
301	7.70377E-06
302	3.85189E-06
303	7.70377E-06
304	1.15557E-05
305	3.85189E-06
309	3.85189E-06
310	3.85189E-06
311	7.70377E-06
314	1.15557E-05
315	7.70377E-06
316	1.15557E-05
317	3.85189E-06
318	3.85189E-06
320	3.85189E-06
322	3.85189E-06
323	3.85189E-06
327	1.54075E-05
329	3.85189E-06
330	3.85189E-06
331	3.85189E-06
332	7.70377E-06
333	3.85189E-06
335	1.15557E-05
338	7.70377E-06
339	7.70377E-06
341	3.85189E-06
342	3.85189E-06
344	7.70377E-06
346	3.85189E-06
350	3.85189E-06
351	3.85189E-06
353	3.85189E-06
354	7.70377E-06
355	3.85189E-06
358	7.70377E-06
360	3.85189E-06
361	3.85189E-06
365	3.85189E-06
367	7.70377E-06
368	3.85189E-06
370	3.85189E-06
372	7.70377E-06
373	3.85189E-06
375	3.85189E-06
376	7.70377E-06
378	3.85189E-06
381	7.70377E-06
382	3.85189E-06
383	3.85189E-06
386	1.15557E-05
388	7.70377E-06
391	7.70377E-06
395	3.85189E-06
397	1.54075E-05
399	3.85189E-06
400	7.70377E-06
402	3.85189E-06
403	3.85189E-06
405	3.85189E-06
408	3.85189E-06
409	3.85189E-06
413	3.85189E-06
414	3.85189E-06
415	1.54075E-05
416	3.85189E-06
417	3.85189E-06
419	7.70377E-06
421	3.85189E-06
425	3.85189E-06
433	3.85189E-06
435	3.85189E-06
437	7.70377E-06
439	3.85189E-06
445	3.85189E-06
446	3.85189E-06
447	3.85189E-06
448	3.85189E-06
449	3.85189E-06
450	3.85189E-06
452	3.85189E-06
458	3.85189E-06
459	3.85189E-06
461	3.85189E-06
464	3.85189E-06
467	3.85189E-06
468	7.70377E-06
472	1.15557E-05
474	3.85189E-06
475	7.70377E-06
476	3.85189E-06
477	3.85189E-06
482	3.85189E-06
483	7.70377E-06
487	3.85189E-06
490	3.85189E-06
492	3.85189E-06
496	3.85189E-06
497	3.85189E-06
500	3.85189E-06
501	3.85189E-06
509	7.70377E-06
520	3.85189E-06
529	3.85189E-06
533	3.85189E-06
538	3.85189E-06
542	3.85189E-06
545	3.85189E-06
547	3.85189E-06
550	3.85189E-06
557	3.85189E-06
558	3.85189E-06
560	3.85189E-06
2613	3.85189E-06
567	3.85189E-06
574	3.85189E-06
575	3.85189E-06
583	3.85189E-06
584	3.85189E-06
585	3.85189E-06
586	3.85189E-06
593	3.85189E-06
594	3.85189E-06
598	3.85189E-06
601	3.85189E-06
603	3.85189E-06
605	3.85189E-06
610	3.85189E-06
613	3.85189E-06
621	3.85189E-06
622	3.85189E-06
623	3.85189E-06
624	3.85189E-06
625	3.85189E-06
638	3.85189E-06
644	3.85189E-06
645	3.85189E-06
655	7.70377E-06
656	3.85189E-06
658	3.85189E-06
660	3.85189E-06
671	3.85189E-06
2723	3.85189E-06
676	3.85189E-06
680	3.85189E-06
681	3.85189E-06
684	3.85189E-06
687	3.85189E-06
689	3.85189E-06
693	3.85189E-06
699	3.85189E-06
704	3.85189E-06
707	3.85189E-06
710	3.85189E-06
715	3.85189E-06
717	3.85189E-06
720	3.85189E-06
723	3.85189E-06
730	3.85189E-06
732	3.85189E-06
742	3.85189E-06
747	3.85189E-06
758	3.85189E-06
768	3.85189E-06
771	3.85189E-06
784	3.85189E-06
790	3.85189E-06
793	3.85189E-06
799	3.85189E-06
802	7.70377E-06
804	7.70377E-06
805	3.85189E-06
806	3.85189E-06
807	3.85189E-06
818	3.85189E-06
828	7.70377E-06
837	3.85189E-06
839	3.85189E-06
855	3.85189E-06
857	3.85189E-06
869	3.85189E-06
886	3.85189E-06
896	3.85189E-06
898	3.85189E-06
902	3.85189E-06
943	3.85189E-06
948	3.85189E-06
962	3.85189E-06
993	3.85189E-06
1056	3.85189E-06
1092	3.85189E-06
1094	3.85189E-06
1146	3.85189E-06
1152	3.85189E-06
1157	3.85189E-06
1160	3.85189E-06
1185	7.70377E-06
1186	3.85189E-06
1229	3.85189E-06
3307	3.85189E-06
1300	3.85189E-06
1319	3.85189E-06
1320	3.85189E-06
3450	3.85189E-06
1451	3.85189E-06
3532	3.85189E-06
3543	3.85189E-06
1561	3.85189E-06
1595	3.85189E-06
1769	3.85189E-06
3905	3.85189E-06
675	3.85189E-06

}{\teamsize}
\begin{tikzpicture}
    \begin{axis}[
    title=Team Size Distribution,
    clip=false,
    width=1.75in,
    height=1.5in,
    ymode=log,
    xmode=log,
    xlabel = \footnotesize{Team Size},
    ylabel = \footnotesize{$p(x)$},
    y label style={at={(axis description cs:-0.23,.5)}},
    ]
    \addplot [only marks, mark=*, blue, mark size=1pt, fill opacity=0.75, draw opacity=0.2]  table [x=Size, y=Number]   {\teamsize};
    \end{axis}
\end{tikzpicture}
    \caption{Like the commit and adoption distributions illustrated in Figs. \ref{fig:commit_dist} and \ref{fig:adopt_dist}, the team size distribution follows a power law.}
    \label{fig:teamsize_dist}
\end{minipage}
\hspace{0.1cm}
\begin{minipage}[t]{0.46\linewidth}
\centering
    \pgfplotstableread{
x	one	two	three	six	ten
1	1.03030303	1.018518519	1.007673132	1.006259919	1
2	1.096969697	1.066137566	1.041006465	1.019080432	1.011111111
3	1.18030303	1.149470899	1.107673132	1.05754197	1.025603865
4	1.264942529	1.242063492	1.183333333	1.122710623	1.050603865
5	1.344109195	1.324461785	1.254166667	1.193223443	1.08578905
6	1.427442529	1.407795118	1.298611111	1.25	1.140079365
7	1.479166667	1.463350674	1.356691919	1.278106222	1.181746032
8	1.5	1.5	1.419191919	1.361439555	1.233010711
9	1.538461538	1.5	1.474747475	1.432868127	1.275338753
10	1.594017094	1.533333333	1.5	1.5	1.29200542
11	1.64957265	1.588888889	1.515151515	1.5	1.305555556
12	1.686868687	1.644444444	1.559011164	1.5	1.305555556
13	1.75036075	1.738506705	1.592344498	1.5	1.333333333
14	1.819805195	1.7515786	1.577192982	1.5	1.388888889
15	1.910714286	1.796023044	1.588888889	1.5	1.444444444
16	1.958333333	1.795179175	1.632791328	1.533333333	1.405797101
17	2	1.893218391	1.688346883	1.557142857	1.305797101
18	2	1.959885057	1.726287263	1.612698413	1.298852657
19	2	2	1.749051491	1.634920635	1.393055556
20	2	2	1.860162602	1.675925926	1.493055556
21	2	2	1.85	1.703703704	1.5
22	2	2	1.916666667	1.814814815	1.523809524
23	2	2	1.916666667	1.903846154	1.54985119
24	2.023809524	2	1.986111111	1.977179487	1.54985119
25	2.023809524	2	1.986111111	1.977179487	1.559375
26	2.09047619	2.007246377	1.986111111	1.99	1.575
27	2.09047619	2.007246377	2	2	1.630555556
28	2.157142857	2.007246377	2	2	1.669025747
29	2.201587302	2.062622549	2	2	1.682914636
30	2.288888889	2.094368581	2	2	1.62735908
31	2.340073529	2.161035247	2	2	1.638888889
32	2.362295752	2.199861974	2	2	1.527777778
33	2.391782614	2.245039019	2	2	1.5638322
34	2.415782414	2.311705686	2	2	1.588013075
35	2.449115747	2.364102564	2	2	1.810235297
36	2.47518444	2.398290598	2	2.03030303	1.903810505
37	2.5	2.442552893	2.077350427	2.03030303	1.796296296
38	2.49122807	2.455373406	2.077350427	2.03030303	1.701058201
39	2.435672515	2.455373406	2.188461538	2	1.738095238
40	2.48112706	2.444444444	2.194444444	2	1.793650794
41	2.545454545	2.444444444	2.254662005	2	1.7858827
42	2.621843434	2.526067123	2.143550894	2	1.738263652
43	2.631944444	2.50153025	2.121756022	2	1.804839662
44	2.576388889	2.50153025	2.144871795	2	1.907845851
45	2.626984127	2.475463127	2.172649573	2	1.788798232
46	2.571428571	2.55	2.245543346	2.066666667	1.80877193
47	2.708769473	2.55	2.273321123	2.066666667	1.7784689
48	2.752364375	2.583333333	2.356654457	2.177777778	1.94114355
49	2.88649136	2.588888889	2.361111111	2.111111111	1.910149398
50	2.915817125	2.672222222	2.413265306	2.177777778	1.940452428
51	2.967460317	2.694444444	2.419801254	2.133333333	1.944444444
52	3	2.638888889	2.414245698	2.133333333	2.006535948
53	3	2.583333333	2.341889483	2.15	2.006535948
54	3	2.611111111	2.390909091	2.183333333	2.006535948
55	2.985915493	2.777777778	2.457575758	2.266666667	2
56	2.985915493	2.861111111	2.533333333	2.35	2
57	2.985915493	2.87037037	2.588888889	2.341954023	2
58	3	2.850762527	2.566666667	2.341954023	2.060606061
59	3	2.906318083	2.6	2.234791714	2.108225108
60	3.025641026	2.847058824	2.522705314	2.253948802	2.074891775
61	2.858974359	2.866666667	2.554451346	2.337282135	2.014285714
62	2.883699634	2.866666667	2.558034799	2.425925926	1.966666667
63	2.913614164	3	2.603583454	2.505291005	2
64	3.08028083	3	2.627392977	2.495487084	2
65	3.055555556	3	2.673809524	2.484593838	2
66	3	3	2.733333333	2.460784314	2
67	3.03030303	3	2.785714286	2.563180828	2
68	3.03030303	3.066666667	2.702380952	2.592592593	2.0625
69	3.113636364	3.066666667	2.719047619	2.592592593	2.0625
70	3.157407407	3.064804469	2.6	2.5	2.0625
71	3.296296296	3.023778828	2.683333333	2.523809524	2
72	3.37962963	3.023778828	2.75	2.49047619	2
73	3.362466125	3.06027306	2.833333333	2.546031746	2
74	3.318815331	3.034632035	2.848232057	2.51965812	2
75	3.235481998	3.034632035	2.755639465	2.552991453	1.962962963
76	3.305892383	3.052380952	2.705639465	2.543359046	1.962962963
77	3.321765399	3.052380952	2.774074074	2.535465633	2.014591011
78	3.371765399	3.076326272	2.825	2.574681319	2.134961381
79	3.411111111	3.066701081	2.891666667	2.614616756	2.151552633
80	3.411111111	3.150034415	2.780555556	2.625074272	2.211035696
81	3.462962963	3.192755762	2.822222222	2.585858586	2.127702363
82	3.376756066	3.261111111	2.888888889	2.5	2.135802469
83	3.265644955	3.288888889	3.014015844	2.666666667	2.024691358
84	3.173793103	3.388888889	3.014015844	2.7	2.024691358
85	3.226666667	3.28989066	3.014015844	2.866666667	2
86	3.44630491	3.257677775	2.857142857	2.81359953	2.035087719
87	3.600834824	3.142162338	2.869047619	2.724710641	2.035087719
88	3.634168157	3.163382789	2.888655462	2.724710641	2.035087719
89	3.552625153	3.160824889	3.114845938	2.777777778	2.166666667
90	3.46031746	3.232887945	3.102941176	3	2.269230769
91	3.543650794	3.282887945	3.083333333	2.962962963	2.435897436
92	3.6498315	3.35589051	3	2.962962963	2.462213225
93	3.760942611	3.375533367	3	2.962962963	2.382871233
94	3.716070816	3.458866701	2.960784314	3	2.327315677
95	3.802041785	3.476190476	2.960784314	2.990196078	2.194939282
96	3.841469448	3.533333333	3.047496025	2.990196078	2.260668762
97	3.969674577	3.584242424	3.253378378	2.990196078	2.205113207
98	3.833872107	3.584242424	3.364489489	2.939632546	2.232226444
99	3.791666667	3.576363636	3.416666667	2.909448819	2.214912281
100	3.583333333	3.5	3.333333333	2.818897638	2.263157895

}{\table}

\begin{tikzpicture}

\pgfplotsset{
every axis legend/.append style={
at={(0.5,-0.30)},
anchor=north
},
}

\begin{axis}[
    legend style={draw=none},
    title = Adoptions by Team Size,
    width=1.75in,
    clip=false,
    xtick = {1,20,40,60,80,100},
    yticklabels={0,100\%,200\%,300\%,400\%,500\%},
    ytick={0,1,2,3,4,5},
    xmin = -1,
    xmax = 100,
    y label style={at={(axis description cs:-0.23,.5)}},
    height=1.25in,
    xlabel = \footnotesize{Commits \# After Adoption},
    ylabel = \footnotesize{median \% change (LOC$_\ell$)},
    legend columns=3
    ]
    \addplot [black, thick]  table [x=x, y=one]   {\table};
    \addplot [blue, thick]  table [x=x, y=two]   {\table};
    \addplot [blue!50!green, thick]  table [x=x, y=three]   {\table};
    \addplot [green, thick]  table [x=x, y=six]   {\table};
    \addplot [orange, thick]  table [x=x, y=ten]   {\table};
    %\draw[ultra thin, dashed] (axis cs:\pgfkeysvalueof{/pgfplots/xmin},0) -- (axis cs:\pgfkeysvalueof{/pgfplots/xmax},0);
    
    \legend{1,2,3-5,6-9,10+}
    
    \node[anchor=west] at (rel axis cs:0.99,0.9) {\scriptsize{1}};
    \node[anchor=west] at (rel axis cs:0.99,0.8) {\scriptsize{2}};
    \node[anchor=west] at (rel axis cs:0.99,0.70) {\scriptsize{3-5}};
    \node[anchor=west] at (rel axis cs:0.99,0.58) {\scriptsize{6-9}};
    \node[anchor=west] at (rel axis cs:0.99,0.43) {\scriptsize{10+}};
    \end{axis}
\end{tikzpicture}
    \caption{Median percentage change in lines of code referencing $\ell$ after adoption.}
    \label{fig:teamsize}
\end{minipage}
\end{figure}

%\begin{figure}[t]
 %   \centering
 %   \include{./figs/teamsize_dist}
%    \caption{Like the commit and adoption distributions illustrated in Figs. \ref{fig:commit_dist} and \ref{fig:adopt_dist}, the team size distribution follows a power law.}
%    \label{fig:teamsize_dist}
%\end{figure}

We calculate a project's team size by counting the number of distinct committers. We observe in  Fig.~\ref{fig:teamsize_dist} that the distribution of team sizes has a power law-like heavy tail wherein 59\% of projects have only a single committer; 24\% and 7\% of projects have two and three distinct committers respectively. Projects with small teams therefore dominate the GitHub community. For the 59\% of projects with a single committer, we do not have to even consider the team dynamics when a library adoption occurs, because only one individual is adopting a new library for use in the project.

Like in the Stack Overflow analysis, we calculate the median growth over the first 100 commits after a library adoption for various team sizes, which we can see in Figure~\ref{fig:teamsize}. We see that smaller teams add more lines of code after the first adoption event than larger teams. A possible explanation for the slower growth of library usage in larger teams is because of perspective differences between two or more committers to a project. Users might feel more comfortable making more commits or experimenting with newly adopted libraries in smaller teams, or if they are working alone, because there are fewer team members to consult with before a commit is made - and also a greater need to ensure that all team members understand the purpose of the commit. Also, more communication might be necessary between teammates before large commits are made, which would appear to cause slower growth rates for bigger teams. It is possible that in larger teams, the first adoption event is more substantial - and then grow more slowly afterwards.

\section{Code Fights}

Fights between committers to a project occur whenever there is a disagreement about how others should structure code, how they should implement features, or any other decision impacting code production. We use this analysis of code fight to understand who wins these arguments by tracking who eventually commits code which eventually stays in the project - or is ultimately removed. Researchers have long analyzed the diffusion and change of information and conventions including work on the adoption of word use in offline textbooks~\cite{perek2014vector}, on Twitter~\cite{goel2016social}, and other domains. The experience of the individuals in the group also plays a key role in what ideas are adopted offline~\cite{krackhardt1997organizational} and in online social systems~\cite{danescu2013no}. 

For example, Sumi et al describe edit wars on Wikipedia where two or more Wikipedia editors constantly edit each other's changes to some article~\cite{sumi2011edit}. Investigators have found fights in collaborative science writing where researchers often use and adapt various \LaTeX~macros and vocabulary. Specifically, based on files obtained from ArXiV, Rotabi et al showed that user experience is a large factor in determining who will win a fight. Less experienced researchers tended to win invisible fights, \ie, fights over \TeX-macros that did not have a high visibility. More experienced researchers, \eg, advisors and senior PIs, tended to win highly visible fights such as fights over the conventions used in the title of the paper~\cite{rotabi2017competition}. In the software development paradigm, norms tend to develop in software teams, to which developers eventually learn and conform~\cite{avery2016externalization}.

In the context of library adoptions in collaborative projects, we informally define a code fight as a series of commits that include back-and-forth additions and deletions of the same code containing a newly adopted $\ell$. For clarity, in the current work we restrict a fight to occur between two committers $u$ and $v$, but we encourage follow-up research that lifts this restriction in future analysis. In this context, a fight occurs when user $v$ removes all (or almost all) of the code that user $u$ committed that references $\ell$. Occasionally, the adopting user $u$ will recommit the original code, which $v$ may then revert. 

A user may add or remove code over a series of contiguous commits, rather than in a single large commit. Therefore, rather than thinking of fights as one commit after another, we model a fight in rounds. A \textit{round} is a series of commits by one user that is uninterrupted by the other user. For example, if $v$ deletes 5 lines of code in commit 2 and then another 6 lines of code in commit 3, then we represent these two commits as a single round with -11 lines deleted.

We formally define a fight as follows. Let $n^{(r)}$ represent the net change in lines of code referencing $\ell$ in round $r$; $r=0$ indicates the round of the adoption event. Also let $n^{\le r}$ be the sum of all lines of code referencing $\ell$ up to and including $r$, \ie, the running total.

%\begin{figure}
%    \centering
%    \include{./figs/probability_of_fight2}
%    \caption{Probability of a fight as a function of team size for various $\epsilon$. An increase in team size is directly correlated with the probability of a fight (Spearman $R^2=0.14$, p$\le0.01$ if $\epsilon$ = 0.1).}
%    \label{fig:probability_of_fight}
%\end{figure}

A \textit{code fight} occurs if there exists any $r$ such that $(1-\epsilon)n^{\le (r-1)} \le n^{\le r}$, where we set $\epsilon$ to represent the percent reduction that must occur, with $\epsilon \in \{.10, .20, .30, .40, .50\}$. Once a fight starts it will continue until there are no more rounds regardless of the size of the change in each round, \ie, further rounds within the same fight do not have to necessarily add or remove $1-\epsilon$ LOC.

The probability of a fight, for various sizes of $\epsilon$ and by project team size, is illustrated in Fig.~\ref{fig:probability_of_fight}. We observe that fights are relatively rare, occurring between 1 and 3 times for every 100,000 commits on average. We also observe that the choice for $\epsilon$ has a limited effect on the probability of a fight. The probability of a fight increases with team size, but with diminishing returns that resemble a Zipfian Distribution. In other words, because there are more interactions in a larger team, it is more likely that a fight will occur.

%\begin{figure}
%    \centering
%    \include{./figs/fight_length}
%    \caption{Fights tend to resolve quickly for various $\epsilon$.}
%    \label{fig:fight_len}
%\end{figure}

Next, we analyze what happens during a two-person fight. Technically speaking, the first round of a fight is the adoption event and the second round of the fight is the removal of at least 100(1 - $\epsilon$) percent lines of the adopter's code. After this point, the two fighters (the adopter and the deleter) may continue with more rounds of back and forth commits.

%\begin{figure}
%    \centering
%    \include{./figs/fights}
%    \caption{When a two-person fight occurs, the adopter $u$ (indicated by odd-numbered %$x$) tends to commit more code than $v$ on average}
%    \label{fig:fights}
%\end{figure}

Despite the dropoff in number of fights, the adopter tends to fight back with more lines of code. In Fig.~\ref{fig:fights} we observe that odd-numbered rounds, corresponding to the adopter, have more net LOC  referencing $\ell$ per round, than the deleter's round that comes afterwards. Also, we see that the larger the original deletion of the code was, the less likely the adoptor is to fight back with lines of code.

\nop{\begin{figure}
    \centering
    \begin{tikzpicture}
\begin{axis}[       
        ybar, 
        bar width=4pt, 
        ymin=0,
        height=2.25in,
        width=3.5in, 
        xtick = {0,1,2,3,4},
        xticklabels = {0.5, 0.4, 0.3, 0.2, 0.1},
        ylabel = {Wins},
]

%50
\addplot+ coordinates
	{
(0, 63650)
(1, 41731)
(2, 33303)
(3, 28832)
(4, 24224)
};

%.60
\addplot+ coordinates
	{
(0, 144344)
(1, 121156)
(2, 111288)
(3, 106469)
(4, 101682)
};

%70
\addplot+ coordinates
	{
(0, 65361)
(1, 47709)
(2, 38677)
(3, 33125)
(4, 27598)
};

\end{axis}
\end{tikzpicture}
\begin{tikzpicture}
    \begin{customlegend}[legend columns=-1,
      legend style={
        draw=none,
        column sep=1ex,
      },
    legend entries={$\epsilon=$, Adopters, Fighters, Non-Adopters}]
    \addlegendimage{blue, only marks, mark=*, fill opacity=0.0, draw opacity=0.0}
    \addlegendimage{only marks, blue, mark=*, fill opacity=0.60, draw opacity=0.9}
    \addlegendimage{only marks, red, mark=*, fill opacity=0.60, draw opacity=0.9}
    \addlegendimage{only marks, pink, mark=*, fill opacity=0.60, draw opacity=0.9}
    \addlegendimage{only marks, black, mark=*, fill opacity=0.60, draw opacity=0.9}
    \addlegendimage{only marks, purple, mark=*, fill opacity=0.60, draw opacity=0.9}
    \end{customlegend}
\end{tikzpicture}
    \caption{We can see that in the majority of fights, the fighters, who are the people who originally delete code from a repository, win the fight as the last committer. Non-adopters, who are parties who were not involved as either the deleter of code or the adopter, win fights as someone who enters a fight later slightly more often than the adopter themselves.}
    \label{fig:fights_win_distribution}
\end{figure}}

We define a fight's winner as the user who was the last commiter referencing $\ell$. In some cases, the adopter may acquiesce to the deleter and allow the code to remain deleted. In other cases, the deleter may allow the adopter's reassertion of the library's addition to stand. By our definition of rounds, it is clear that the deleter wins approximately 90\% of the fights because the adopter only fights back 10\% of the time. This shows that it is relatively rare for users to counter and re-add code. Perhaps in some cases, the ``fight'' was not contentious and the result of a mutual agreement to remove a library. Further research is needed to find out how many of these fights are a result of team friction.

What role does experience play in winning a fight? To answer this question, we must first ask how to best define experience. Two options are to count 1) the number of commits of a user (in any project) or 2) the time elapsed since the users first commit (in any project). Although these two options obtain similar results, the current work maintains the standard set by prior studies~\cite{rotabi2017competition} and therefore defines experience as the time since the user's first commit (in any project). We observe in Fig.~\ref{fig:fights_experience}, which plots only results from $\epsilon=0.1$, that the more experienced committer wins the fights between 70\% and 80\% of the time. Results from alternative $\epsilon$ values were nearly identical to $\epsilon = 0.1$ and are omitted for clarity and brevity. The experience difference groupings were selected so that each contained a similar number of fights. Interestingly, the more experienced users have about a 75\% win probability even when the experience differences are less than a week or even a day (not illustrated). This suggests that even slightly more familiarity with the project (perhaps an indication of project leadership) results in the more experienced user winning the fight. It appears that it is more common for new people working on a team to be subservient to the person who has been working in the codebase the most.

\begin{figure*}[t]
\begin{minipage}[t]{0.33\linewidth}
\centering
    \pgfplotstableread{
x   n  e  s  six  f
2	1.46695E-06	1.25771E-06	1.18614E-06	1.15547E-06	1.12674E-06
3	2.12572E-06	1.88679E-06	1.79096E-06	1.75219E-06	1.71234E-06
4	2.48671E-06	2.2474E-06	2.14087E-06	2.10286E-06	2.07127E-06
5	2.76293E-06	2.5583E-06	2.47832E-06	2.42535E-06	2.39419E-06
6	2.88998E-06	2.71213E-06	2.64715E-06	2.60611E-06	2.56849E-06
7	2.97757E-06	2.82031E-06	2.74555E-06	2.71977E-06	2.70946E-06
8	3.11145E-06	2.95281E-06	2.86268E-06	2.81941E-06	2.77975E-06
9	3.12647E-06	2.99144E-06	2.93431E-06	2.88757E-06	2.8668E-06
10	3.1334E-06	3.0446E-06	2.98117E-06	2.94311E-06	2.93043E-06
11	3.13779E-06	3.02515E-06	2.97688E-06	2.96079E-06	2.93665E-06
12	3.24635E-06	3.14101E-06	3.09313E-06	3.04525E-06	3.0261E-06
13	3.22779E-06	3.07963E-06	3.05846E-06	3.02672E-06	3.00555E-06
14	3.25064E-06	3.12317E-06	3.07217E-06	3.02118E-06	2.99569E-06
15	3.35368E-06	3.28825E-06	3.28825E-06	3.25553E-06	3.23917E-06
16	3.33138E-06	3.2124E-06	3.2124E-06	3.2124E-06	3.2124E-06
17	3.28711E-06	3.24145E-06	3.17297E-06	3.12732E-06	3.12732E-06
18	3.47956E-06	3.43254E-06	3.3385E-06	3.26797E-06	3.22095E-06
19	3.20773E-06	3.13075E-06	3.13075E-06	3.07942E-06	3.05376E-06
20	3.29686E-06	3.26689E-06	3.26689E-06	3.26689E-06	3.20695E-06
21	3.20452E-06	3.06018E-06	3.00244E-06	3.00244E-06	3.00244E-06
22	3.293E-06	3.293E-06	3.293E-06	3.26133E-06	3.26133E-06
23	3.23162E-06	3.05041E-06	3.0202E-06	2.99E-06	2.99E-06
24	3.2588E-06	3.2588E-06	3.2588E-06	3.21954E-06	3.21954E-06
25	3.1872E-06	3.14101E-06	3.09482E-06	3.09482E-06	3.04863E-06
26	3.52563E-06	3.52563E-06	3.52563E-06	3.52563E-06	3.48556E-06
27	3.56523E-06	3.5201E-06	3.5201E-06	3.5201E-06	3.47497E-06
28	3.47323E-06	3.32222E-06	3.22155E-06	3.22155E-06	3.22155E-06
29	3.42857E-06	3.42857E-06	3.37327E-06	3.37327E-06	3.26267E-06
30	3.30305E-06	3.24073E-06	3.24073E-06	3.24073E-06	3.24073E-06
31	3.43548E-06	3.43548E-06	3.39087E-06	3.39087E-06	3.39087E-06
32	3.32222E-06	3.24672E-06	3.17121E-06	3.17121E-06	3.17121E-06
33	2.99461E-06	2.92806E-06	2.92806E-06	2.92806E-06	2.92806E-06
34	3.293E-06	3.22967E-06	3.22967E-06	3.16634E-06	3.16634E-06
35	3.24672E-06	3.0202E-06	3.0202E-06	3.0202E-06	2.9447E-06
36	3.41134E-06	3.41134E-06	3.41134E-06	3.28261E-06	3.28261E-06
37	3.71593E-06	3.64582E-06	3.64582E-06	3.64582E-06	3.64582E-06
38	3.23339E-06	3.23339E-06	3.23339E-06	3.00244E-06	3.00244E-06
39	3.01318E-06	3.01318E-06	3.01318E-06	2.92187E-06	2.92187E-06
40	2.81667E-06	2.81667E-06	2.81667E-06	2.81667E-06	2.81667E-06
}{\table}

\begin{tikzpicture}

    \begin{axis}[
    title = {Probability of Fight},
    height=1.5in,
    width=2.5in,
    xlabel = {Team Size},
    ylabel = {$p(x)$},
    xmax = 20.5,
    xmin = 1.5,
    y label style={at={(axis description cs:-0.05,.5)}},
    ]
    \addplot+ [only marks, mark=*, blue, mark size=2pt, fill opacity=0.60, draw opacity=0.9]  table [x=x, y=f] {\table};
    \addplot+ [only marks, mark=*, red, mark size=2pt, fill opacity=0.60, draw opacity=0.9]  table [x=x, y=six] {\table};
    \addplot+ [only marks, mark=*, pink, mark size=2pt, fill opacity=0.60, draw opacity=0.9]  table [x=x, y=s] {\table};
    \addplot+ [only marks, mark=*, black, mark size=2pt, fill opacity=0.60, draw opacity=0.9]  table [x=x, y=e] {\table};
    \addplot+ [only marks, mark=*, purple, mark size=2pt, fill opacity=0.60, draw opacity=0.9]  table [x=x, y=n] {\table};
    
    \end{axis}

\end{tikzpicture}
\begin{tikzpicture}[scale=0.6]
    \begin{customlegend}[legend columns=-1,
      legend style={
        draw=none,
        column sep=1ex,
      },
    legend entries={$\epsilon=$, 0.5, 0.4, 0.3, 0.2, 0.1}]
    \addlegendimage{blue, only marks, mark=*, fill opacity=0.0, draw opacity=0.0}
    \addlegendimage{only marks, blue, mark=*, fill opacity=0.60, draw opacity=0.9}
    \addlegendimage{only marks, red, mark=*, fill opacity=0.60, draw opacity=0.9}
    \addlegendimage{only marks, pink, mark=*, fill opacity=0.60, draw opacity=0.9}
    \addlegendimage{only marks, black, mark=*, fill opacity=0.60, draw opacity=0.9}
    \addlegendimage{only marks, purple, mark=*, fill opacity=0.60, draw opacity=0.9}
    \end{customlegend}
\end{tikzpicture}
    \caption{Probability of a fight as a function of team size for various $\epsilon$. An increase in team size is directly correlated with the probability of a fight (Spearman $R^2=0.14$, p$\le0.01$ if $\epsilon$ = 0.1).}
    \label{fig:probability_of_fight}
\end{minipage}
\hspace{0.2cm}
\begin{minipage}[t]{0.33\linewidth}
\centering
    \begin{tikzpicture}
\begin{axis}[       
        ybar, 
        bar width=2pt, 
        xmax=7.5,
        xmin=1.5,
        ymin=0,
        %ymode=log,
        height=1.5in,
        title=Fight Size,
        width=2.5in,
        y label style={at={(axis description cs:-0.05,.5)}},
        xlabel = {Round},
        xtick = {2,3,4,5,6,7},
        xticklabels = {$3_u$,$4_v$,$5_u$,$6_v$,$7_u$,$8_v$},
        ylabel = {Net LOC$_\ell$ per fight},
]

%50
\addplot+ coordinates
	{
(0, 24.58701664403882)
(1, -20.84974480816614)
(2, 2.193977221300347)
(3, 0.5654824759893398)
(4, 0.5608689093377583)
(5, 0.12530799014431537)
(6, 0.19011791622668076)
(7, 0.06275456328254639)
(8, 0.05572735958163624)
(9, 0.04858701664403882)
};

%.60
\addplot+ coordinates
	{
(0, 28.49826319774567)
(1, -25.323942892380707)
(2, 1.38937935664161)
(3, 0.4690771140200923)
(4, 0.5344121021942252)
(5, 0.0831557501814860)
(6, 0.156663466267007)
(7, 0.04325223052244573)
(8, 0.03600839909764341)
(9, 0.03197277318533436)
};

%70
\addplot+ coordinates
	{
(0, 27.221312110131212)
(1, -25.13420520542052)
(2, 1.2710174231017424)
(3, 0.3865691546569155)
(4, 0.2876575607657561)
(5, 0.05107334910733491)
(6, 0.10212949021294902)
(7, 0.034699935469993545)
(8, 0.02906431490643149)
(9, 0.025837814583781458)
};

%80
\addplot+ coordinates
	{
(0, 24.095782216914557)
(1, -23.079352585676546)
(2, 1.0595574132377954)
(3, 0.3807404190641624)
(4, 0.3277240979987696)
(5, 0.04141786993811747)
(6, 0.03759092389534253)
(7, 0.029258495277385735)
(8, 0.024481598089241125)
(9, 0.01049469836789346)
};

%90
\addplot+ coordinates
	{
(0, 18.120241398049366)
(1, -17.92301385273956)
(2, 0.9004090038040214)
(3, 0.3845207791093442)
(4, 0.29087892915367675)
(5, 0.020593198524154106)
(6, 0.016922651564987748)
(7, 0.01855294644814995)
(8, 0.00824681329786727)
(9, 0.0021546587344716795)
};

\end{axis}
\end{tikzpicture}
\begin{tikzpicture}[scale=0.4]
    \begin{customlegend}[legend columns=-1,
      legend style={
        draw=none,
        column sep=0.5ex,
      },
    legend entries={$\epsilon=$, 0.5, 0.4, 0.3, 0.2, 0.1}]
    \addlegendimage{ybar,ybar legend, fill opacity=0.6, draw opacity=0.0}
    \addlegendimage{ybar,ybar legend, fill=blue, fill opacity=0.40, draw opacity=0.9}
    \addlegendimage{ybar,ybar legend, fill=red, fill opacity=0.40, draw opacity=0.9}
    \addlegendimage{ybar,ybar legend, fill=brown, fill opacity=0.40, draw opacity=0.9}
    \addlegendimage{ybar,ybar legend, fill=black, fill opacity=0.40, draw opacity=0.9}
    \addlegendimage{ybar,ybar legend, fill=purple, fill opacity=0.90, draw opacity=0.9}
    \end{customlegend}
\end{tikzpicture}
    \caption{When a two-person fight occurs, the adopter $u$ (indicated by odd-numbered $x$) tends to commit more code than $v$ on average}
    \label{fig:fights}
\end{minipage}
\hspace{0.2cm}
\begin{minipage}[t]{0.25\linewidth}
\centering
    \pgfplotstableread{
x	nintety
0	0.727379553
1	0.781195589
2	0.759059745
3	0.718345384
4	0.793791574
}{\winprob}

\begin{tikzpicture}

    \begin{axis}[
    height=1.5in,
    width=2.25in,
    title=Fight Winners by Experience,
    xtick = {0,1,2,3,4},
    xticklabels = {{[0,30d)}, {[30d,6m)}, {[6m,1y)}, {[1y,4y}), {[4y,$\infty$})},
    ylabel = {Prob. exp. wins},
    xlabel = {Experience Difference},
    ymin= 0.0,
    ymax=1.0,
    y label style={at={(axis description cs:-0.10,.5)}},
    ]
    \addplot+ []  table [x=x, y=nintety]   {\winprob};
    
    \end{axis}

\end{tikzpicture}
    \caption{The more experienced user is more likely to win a code fight regardless of the experience gap.}
    \label{fig:fights_experience}
\end{minipage}
\end{figure*}

%\begin{figure}
%    \centering
%    \include{./figs/Fight_experience}
%    \caption{The more experienced user is more likely to win a code fight regardless of %the experience gap.}
%    \label{fig:fights_experience}
%\end{figure}

Finally, we ask: which libraries are the most fought over? To answer this question, we counted the occurrences of $\ell$ within each two-person fight and normalized by the number of times $\ell$ was adopted. Here we set $\epsilon=0.1$, but results for other $\epsilon$ were similar. Table~\ref{tab:libs} shows the top eight libraries involved in the most fights. 

\begin{table}[t]
    \centering
    {\small
    \begin{tabular}{c|c}
    Library & Prob in Fight \\ \hline
    pdb & 12.4 \\
    pprint & 11.3 \\
    telnetlib & 10.5 \\
    syslog & 10.3 \\
    distutils & 9.1 \\
    glob & 9.0 \\
    poplib & 8.4 \\
    imp & 7.8 \\
    \end{tabular}
    }
    \caption{Libraries most likely to be involved in a fight.}
    \label{tab:libs}
\end{table}

We observe that common debugging libraries \texttt{pdb}, \texttt{pprint}, and \texttt{syslog} comprise three of the top four most common causes of fights. In these instances, we assume that adopters might be implementing or testing some functionality, they use the debugging libraries, and commit changes with the debugging code still implemented. This code then is then removed by another team member thereby instigating a fight.

It is not surprising to see the \texttt{distutils} library counted among the top fight starters. This particular library is used to generate source code distributions \ie, code releases, but it is strongly encouraged that users use the \texttt{setuptools} library instead. So most cases importing \texttt{disutils} is likely an error, and the usage of that library was deleted by others who are updating their code base, which appears in our analysis as a code ``fight.'' However, we can suppose that for some teams, the discussion of using \texttt{setuptools} in favor of \texttt{disutils} is indeed a fight, especially if a team member is reluctant to switch libraries.

\nop{
\begin{table}[t]
    \centering
    \begin{tabular}{c|c}
        Fight Threshold  & \% Median Net Commit Size \\ \hline
0.5 & 24.58 \\
0.4 & 28.49 \\
0.3 & 27.22 \\
0.2 & 24.09 \\
0.1 & 18.12 \\
    \end{tabular}
    \caption{Various values for median net commit lines before a fight occurs, in which the next commit results in net lines of code that are below the set fight threshold.}
    \label{tab:fight_before}
\end{table}
}

\section{Discussion}

Let's return to our original questions and summarize our  results based on our findings.

\paragraph{What does it look like when a team adopts a library for the first time?}

In Fig.~\ref{fig:adopt_per_commit} we observe that library adoptions tend to happen early in a project's history. Hence, the probability of a new library being adopted later is lower.  We can expect that it is difficult to adopt a new library once a project has matured. Perhaps this is because new libraries may introduce instability into a repository, or because the primary innovation within a project occurs early on in its lifespan. Further research is needed to understand this more clearly. 

While Fig.~\ref{fig:adopt_per_commit} shows us that these adoptions happen early on, it would also be interesting to approach this problem from the perspective of \textit{percent} of project repository history. We see that the commit distribution follows a power law distribution, and many projects have few commits. Therefore, while we measure the commits in a sequential order in Fig.~\ref{fig:adopt_per_commit}, further research should be done to see when library adoptions occur as a measure of project completeness. Early in a project history, do we see that it takes more time for a library to be adopted? Or do we see that a team takes longer to adopt a new project after they have been working using established methods? Further research needs to be done to answer these questions. Despite these questions, we can see that library adoptions tend to be events that happen in the first few commits - though further research can help us understand when they occur in relation to the length of a project. That would give us another dimension to analyze this problem.

\paragraph{Are commits containing new libraries more likely to have deletions than other types of commits?}

Once an adoption has occurred, we track how long it takes for library usage to become stable, or adopted, within the project by examining how many additions and deletions occur in the commits after a library is first used. In Fig.~\ref{fig:afteradoptionteamsize}, we observe that activity involving a newly adopted library is relatively high after a commit occurs.  Over time, the number of lines of code referencing the adopted library stabilizes, which indicates that the library has been fully adopted and incorporated into the project. We can safely conclude that users tend to write most lines of code that involve a newly adopted library relatively soon (within 10-15 commits) after library adoption. However, in many instances we also find that some libraries never stabilize and end up being deleted.

While we currently measure when adoptions occur, it would also be interesting to see when libraries are completely deleted from project repositories. When would those deletions occur? Would we find that libraries are more likely to be deleted early on in a project repository history, or later? Would these deletions follow a similar distribution over time as library adoptions? To answer these questions, we need more research from the perspective of library deletions. As with Research Question 1, while we are currently asking this question from the perspective of commit number, we can also use the percentage of project completeness to understand when these deletions are occurring over the lifespan of a project. An analysis such as this would not skew the averages towards smaller projects since the project timeline would be measured from the perspective of percentage of a project completed, instead of absolute commit number.

Additionally, we could ask further questions about the coding history of users that are on the project. To add to this work, we could track the libraries used by the individuals who are contributing to these projects. Other work as shown that user activity rates are higher soon after an adoption event~\cite{krohn2019library}. Tracking user history across GitHub repositories could give us more information about what occurs in an adoption event, and help us learn more about how individuals learn and retain new information.

\paragraph{Do the answers to these questions vary by library type, team size, or the amount of information available on Stack Overflow?}

When team sizes are larger, the lines of library code do not grow as quickly relative to the first adoption commit as they do when team sizes are smaller. This may be because larger team projects require more communication and planning and are therefore less agile than small teams or individual projects. As mentioned previously, this is one of the drawbacks of using lines of code as a measurement tool, since it could be argued that the code committed by larger teams is more valuable than the lines of code committed by smaller teams. Additionally, this work only looked at public software repositories. Perhaps an analysis of private software repositories would lead to different conclusions about how various team sizes work together - because the interaction between teammates who know each other personally may be different than those who do not, or may be operating in a public environment.

Further in-depth research is necessary here. While we defined `large' team sizes as those with 10+ members, it would be interesting to see how teams whose members number in the hundreds or thousands would differ. We would expect these teams to have different adoption behavior. Also, a team size and lifespan analysis would provide unique insights into how long large team repositories survive versus smaller teams. What is the distribution of smaller team projects' lifespans compared to larger teams? 

In addition to team size, we showed that the number of times a library appears in Stack Overflow is highly correlated with the number of adoptions in our data set. Further questions about StackOverflow still need to be answered. In particular, more questions need to be answered about which way StackOverflow and GitHub grow - does StackOverflow influence GitHub, does GitHub influence StackOverflow, or is there information flowing in both directions? How do individuals find new libraries on StackOverflow? What is the distribution of attention given to different StackOverflow questions? Each of these questions requires further analysis as we attempt to understand how groups utilize outside resources to learn about new information.

\paragraph{What does it look like when team members fight over new library usage?}

When working on a team, there is bound to be conflict. Different team members have various opinions about which library is best to use in a repository. The probability of these fights occurring increases with team size, as shown in Fig.~\ref{fig:probability_of_fight}. The winner of these fights tends to be more experienced as shown in Fig.~\ref{fig:fights_experience}.

While we have done an analysis of team size, length of experience, and libraries used in fights, there are still questions to be answered regarding fights. Further research directions could include analyzing when fights occur in a project history. It would be interesting to see if fights are something that occur early or late in a project repository history. Additionally, this research was focused solely on fights between two individuals. Further analysis needs to be done on larger team sizes. Would an analysis of large team fights yield groups of people who are fighting for control on a project? Would larger teams result in more or less fights? 

It makes intuitive sense that more experienced teammates would win GitHub fights, since they have more seniority in a project. However, it would also be interesting to investigate the projects in which the individual with less project experience wins the fight. What characteristics do these new, inexperienced team members have that causes them to win code fights? Is there another interesting quality that results in them having more influence over their teammates? More research is needed to answer these questions.

Additionally, in future work we could to uncover who is bringing new libraries into a project. We have seen that more experienced programmers tend to win these code fights. However, we could learn more information about how teammates work together by uncovering which team members are the first to bring in a new library. Would the teammates who have worked longer on the project be the ones to try new approaches with other libraries? Or would new teammates be the ones to suggest new ideas? Additionally, is there a divide between who is using new libraries and the popularity of that library on StackOverflow? We could hypothesize that common libraries, such as \texttt{os}, would be used for the first time equally by both experienced and inexperienced programmers, but a more specialized library like \texttt{numpy} would only be used by veteran team members. Since we have suggested libraries such as \texttt{pandas} have steep learning curves due to the number of posts on StackOverflow about that particular library, we could also track how team members fight over how to use complicated libraries, and if those libraries have a longer time to adoption than others.

\paragraph{Implications}

There are some important caveats to the findings presented in the present work. We were only able to crawl 13\% of all public Python GitHub projects; even if we could obtain all Python projects, these public projects only represent a subset of all Git projects. Therefore, we must temper conclusions to represent a case study of this domain and we caution the reader against drawing broad conclusions about user behavior outside of Python projects from public GitHub repositories. We can hypothesize that private software repositories are written by teams that have higher interpersonal relationships, and library adoption will appear to be different in these groups. Therefore, our conclusions cannot be generalized to all GitHub projects, though they provide a good overview of how libraries are adopted in public GitHub repositories. In particular, fights might look different in private repositories because we can assume that there would be some sort of relationship between the committers in those repositories, which might effect who wins those code fights. Additionally, we might find that people who post only in private repositories might be very different from those who have public GitHub accounts. Understanding the differences between these types of users would yield more interesting takeaways about how individuals work together in teams, as we analyze the difference between people who are more guarded of their work to those who are more open.

From our work, we found some interesting takeaways. We see that the number of commits and adoptions per project, along with team size, follow a power law distribution. We can conclude that power law distributions are common as we understand social behavior - there are many people who contribute little, and only a few that contribute in large amounts. We found positive correlations between the number of times libraries appear in StackOverflow and GitHub. We discovered that popular libraries on StackOverflow have faster rates of adoption for projects in Git. Additionally, smaller teams are more agile and can grow more quickly than larger teams, when productivity is measured as a function of median percentage growth. We also find that code fights are rare, but when they occur, they tend to be won by more experienced coders, and involve libraries which are used for debugging purposes. We can therefore conclude that the availability and proliferation of online resources has helped improve productivity for programmer, and that team characteristics, including size and seniority, have interesting implications for team dynamics. We can only expect resources such as StackOverflow and GitHub to continue growing as they attract new users.

However, just analyzing commit histories of varied team sizes do not tell the whole story. While we see in Figure~\ref{fig:teamsize} that larger teams do not grow as quickly, further research needs to be done to conclude if these larger teams are actually more efficient. These smaller teams may be `moving fast and breaking things,' while larger teams could be more cautious in their execution due to the fact that larger teams by definition have more interpersonal relationships which need to be managed. It could be that larger team sizes have a higher percentage of their code that makes its way into the production version, while the multiple, minor commits that smaller teams make might be junk code that winds up being deleted. Therefore, this research needs to be continued to investigate this problem of team productivity from many different angles. This begs the quesiton of what `productivity' means, which could fuel many different research questions.

While this work has focused on the library-centric approach to understanding adoptions, more work needs to be done to understand how individuals work to adopt information. Epidemiological models have been used to understand how information spreads across a social network~\cite{zhao2013sir}, with a SIR model (suspectible, infected, recovered) being used to mimic one's potential to become `infected' (or viewing a post) and `infecting' it (by sharing it). In these models, individuals have varying levels of susceptibility. Further research could apply these models to GitHub users. Do we see that there are some individuals which have high susceptibility rates, which could enable them to adopt to new libraries more quickly? Also are there some users who are more `infecting' than others, and when they are present on a project, their introduced libraries are adopted more quickly than those introduced by individuals who are less `good' at spreading `infections.' This could be an interesting research topic that could attempt to apply epidemiological models to GitHub projects, where a library could be considered a virus. This type of approach could also help us create a model where we find highly influential GitHub users that are highly successful in implementing new libraries in varied projects. We could also view new libraries as a `virus' that spread throughout GitHub - and track the users that are instrumental in helping them spread.

From this work, we can conclude that the proliferation of online coding resources such as GitHub and StackOverflow have been a positive development for programmers who wish to learn how to use new libraries to accomplish their coding tasks. Observing the growth of individuals who use StackOverflow and GitHub over time shows that the platforms have had tremendous growth over time for those hoping to learn to program. We expect continued high growth rates for both programs.

\paragraph{Future Work}

We uncovered some interesting findings, but ultimately end up with more questions than conclusive answers. We encourage the community to explore specific questions raised by our results using the methodology developed in the present work. Specifically, we encourage further probing into how Stack Overflow contributes to the growth of library adoption and popularity. We have uncovered patterns that exist when team members fight over library adoption, and we look forward to further research which investigates code fights at an even greater depth.

We present several facets of analysis, but due to the complexity of varying size of teams and GitHub repositories, there exists virtually limitless possibilities in exploring the data. Since the commit and adoption distribution follow a power law, it might be interesting to investigate what occurs only in projects where these values are very large. This paper attempted to account for this variety of team sizes, but more focused work on investigating either small or large team sizes would yield very interesting results. Additionally, tracking the growth of team sizes could yield some very interesting research.

While this work ignores \textit{when} commits occurred as a matter of date and time, gathering time stamps of commits might also yield interesting research. There are several questions that could be answered by analyzing time stamps, such as which time of year projects are more productive (students trying to finish semester projects? Teams being more productive at the end of the month?).

Additionally, this work does not attempt to track users across projects. Further research could be done to discover if a person's prior programming experience results in lower time to adoption, or if sufficiently complex libraries remain hard even for experienced programmers. Some of these questions have been answered by Krohn et. al~\cite{krohn2019library}. Another interesting research topic would be to track how quickly libraries are adopted as a function of time spent in the project, since many projects have short lifespans. Further research could attempt to find if libraries that are adopted later in a project's repository history will be adopted more quickly, as teams learn how to work together more cohesively.

\paragraph{Conclusion}

We have presented an analysis of the dynamics of library adoptions in Python projects pushed to GitHub. We find that when teams attempt to learn new information together, it can be challenging to apply these new concepts and there is often a learning curve needed before the new information can fully stabilize within the group project. We further find that that even though learning curves are unavoidable, it helps to have teammates and other online resources that can guide groups towards learning how to adopt new information.  When conflict arises, the more experienced team members usually end up winning disagreements when we track whose code ends up in the final version. Through this work, we confirmed that learning new information together can be a difficult process, and that many of the statistics surrounding GitHub projects follow power law distributions (including team size, commits, and adoptions). This work provides a superficial glimpse into an analysis of teamwork on GitHub, though there are still more in-depth questions to be answered to uncover more information about more complex interactions.

%\newpage

%\section*{Acknowledgement}

% Generated by IEEEtran.bst, version: 1.14 (2015/08/26)

\end{document}